\documentclass[usenatbib]{mnras}

\usepackage[T1]{fontenc}
\usepackage{caption}
\usepackage{subcaption}
\usepackage{graphicx}	% Including figure files
\usepackage{amsmath}	% Advanced maths commands
\usepackage{siunitx}
\usepackage{newtxtext,newtxmath}
\usepackage{xcolor}
\title[A late--Universe approach to modern cosmology]{An analytical late--Universe approach to the weaving of modern cosmology}

\author[Cogato, Moresco, Amati \& Cimatti]{
Fabrizio Cogato$^{1,2}$\thanks{E-mail: fabrizio.cogato@inaf.it},
Michele Moresco$^{1,2}$\thanks{E-mail: michele.moresco@unibo.it},
Lorenzo Amati$^{2}$\thanks{E-mail: lorenzo.amati@inaf.it}
and Andrea Cimatti$^{3,2}$\thanks{E-mail: a.cimatti@unibo.it}
\\
$^{1}$ Dipartimento di Fisica e Astronomia ``Augusto Righi'' - Università di Bologna, via Piero Gobetti 93/2, I-40129 Bologna, Italy\\
$^{2}$ INAF - Osservatorio di Astrofisica e Scienza dello Spazio di Bologna, via Piero Gobetti 93/3, I-40129 Bologna, Italy\\
$^{3}$ Dipartimento di Fisica e Astronomia ``Augusto Righi'', Universit\`a di Bologna, Viale Berti Pichat 6/2, I-40127, Bologna, Italy
}

\date{Accepted 2023 November 14. Received 2023 November 14; in original form 2023 September 6}

\pubyear{2023}

\begin{document}
\label{firstpage}
\pagerange{\pageref{firstpage}--\pageref{lastpage}}
\maketitle

\begin{abstract}
Combining cosmological probes has consolidated the standard cosmological model with percent precision, but some tensions have recently emerged when certain parameters are estimated from the local or primordial Universe. The origin of this behaviour is still under debate, however, it is crucial to study as many probes as possible to cross--check the results with independent methods and provide additional pieces of information to the cosmological puzzle.
In this work, by combining several late--Universe probes (0$<z<$10), namely, Type Ia SuperNovae, Baryon Acoustic Oscillations, Cosmic Chronometers and Gamma--Ray Bursts, we aim to derive cosmological constraints independently of local or early--Universe anchors. 
To test the standard cosmological model and its various extensions, considering an evolving Dark Energy Equation of State and the curvature as a free parameter, we analyse each probe individually and all their possible permutations.
Assuming a flat $\Lambda$CDM model, the full combination of probes provides $H_0=67.2^{+3.4}_{-3.2}$ km s$^{-1}$ Mpc$^{-1}$ and $\Omega_m=0.325\pm0.015$ (68$\%$ C.L.). Considering a flat $w$CDM model, we measure $w_0=-0.91^{+0.07}_{-0.08}$ (68$\%$ C.L.), while by relaxing the flatness assumption ($\Lambda$CDM model, 95$\%$ C.L.) we obtain $\Omega_k=0.125^{+0.167}_{-0.165}$.
Finally, we analytically characterize the degeneracy directions and the relative orientation of the probes' contours. By calculating the Figure--of--Merit, we quantify the synergies among independent methods, estimate the constraining power of each probe and identify which provides the best contribution to the inference process.
Pending the new cosmological surveys, this study confirms the exigency for new emerging probes in the landscape of modern cosmology.

\end{abstract}


%%%%%%%%%%%%%%%%%%%%%%%%%%%%%%%%%%%%%%%%%%%%%%%%%%

%%%%%%%%%%%%%%%%% BODY OF PAPER %%%%%%%%%%%%%%%%%%

\section{Introduction}
At the beginning of the twenties of the third millennium, the discovery of the accelerated expansion of the Universe radically changed our understanding of its origin and evolution. 
To date, the physical nature of the energy driving this expansion -- commonly called Dark Energy -- and of most of the matter components in the Universe -- referred as Cold Dark Matter (CDM) -- still animate the debate within the scientific community.

Despite the fact that we are still unsure of the exact nature of these components, with the advanced technologies at our disposal we are able nowadays to precisely measure their effects on the observable Universe. 
For example, it has been observed that the total energy budget of the cosmos roughly matches its critical value, so that the flatness of the Universe is inferred with an extremely high level of precision \citep{planck18}. 
The properties of the cosmological fluid responsible for the current accelerated expansion can also be measured, although to accurately study its possible temporal evolution we have to wait until more advanced surveys start to observe the sky, such as \textit{Euclid} \citep{euclid}, \textit{Vera Rubin Observatory} \citep{lsst} and \textit{Nancy Grace Roman Space Telescope} \citep{wfirst}.

Moreover, another new fact emerged recently in the already complicated weaving of modern cosmology: the present--day expansion velocity (the Hubble constant, $H_0$) presented a significant discrepancy when measured with independent probes, leading to the well--known \textit{Hubble tension} \citep{hubbletension}.
Currently, this tension is mostly driven by the difference between the $H_0$ estimated with Cepheids and Type Ia SuperNovae \citep[SNe, $H_0=73.04\pm1.04$ km s$^{-1}$ Mpc$^{-1}$,][]{shoes} and with the analysis of Cosmic Microwave Background radiation from the ESA mission Planck \citep[CMB, $H_0=67.36\pm0.54$ km s$^{-1}$ Mpc$^{-1}$,][]{planck18}, but some evidence suggest a dichotomy between late--Universe and early--Universe cosmological probes \citep{abdalla22}.
To solve this tension, several alternative models have been proposed \citep{divalentino} that, however, need a deeper and more detailed comparison with observations \citep{nils22}.
Of course, before abandoning the standard cosmological model, a detailed assessment of the possible systematic effects is necessary. To address this delicate topic, one of the most studied methods is definitively the statistical combination of cosmological probes.

The combination of probes is hardly a new concept in cosmology. Since an additional cosmological component was measured by \citet{riess98} and \citet{perlmutter}, this method has been widely used to increase the precision of cosmological parameters inference \citep{bridle01, linder06, davis07, lamp10, suzuki12, huterer18, nilsverde22, brieden23}.
Essentially, by combining the likelihoods obtained from independent probes, it is possible to exploit the different constraint powers of each observable to alleviate the degeneracy among cosmological parameters and thus improve our capability to investigate a wide range of models.
Currently, only a few probes have been so extensively studied that they now represent the standard for any cosmological analysis, including Cosmic Microwave Background, Type Ia SuperNovae and Baryon Acoustic Oscillations. While the former samples the early stages of the cosmos ($z\sim1100$), the others are observed at low--redshift up to $z\sim2.5$.

In this paper, Type Ia SuperNovae and Baryon Acoustic Oscillations are analysed in synergy with two of the new emerging methods recently developed in the context of late--Universe cosmological probes \citep{morescorev22}. On the one hand, Cosmic Chronometers provide independent measurements of the Hubble parameter and, thus, are able to probe the cosmic expansion history up to $z\sim2$. On the other, Gamma--Ray Bursts offer the opportunity to extract the cosmological signal up to $z\sim10$. 
Hence, the scope of this work is to provide a detailed study of these late--Universe probes, analysing the most recent data collections and combining them to obtain new and precise cosmological constraints that are completely independent of the standard approach in modern cosmology, namely, the Cosmic Microwave Background \citep{planck18} or the three--rung distance ladder \citep{shoes_pantheon+}.

Our approach essentially consists of a probe--by--probe combination through which we aim to monitor the potential systematic effects and compare the constraining power of each probe. 
In fact, our general purpose is to consolidate and extend the constraints from these late--Universe probes (especially for the emerging ones), ensure the robustness of the combination technique, and finally obtain reliable estimates of cosmological parameters that are noteworthy in the modern cosmology landscape.

Finally, we seek to define a mathematical framework to properly assess the complementarity between different cosmological probes and effectively exploit the synergies among their constraint powers.
In practice, starting from the idea developed by \citet{linder06}, the peculiar degeneracy affecting each probe is evaluated by looking at the confidence contours' orientation on the parameter planes. From the largest eigenvalue of the covariance matrix associated with these contours, we trivially calculate the relative orientation of different probes and, by means of the Figure--of--Merit defined by \citet{albrecht06} and \citet{yunwangFoM}, evaluate the effects of probes combination. 
From this perspective, we derive some useful insights on the most effective combination to improve the inferential process as a function of the cosmological model.

To summarize, the physics explored by an experiment is strictly defined by the orientation of confidence contours in a particular parameter space. Hence, by fixing the parameter space and the probes to constrain it, the falsifiable models are closely related to the degeneracy directions affecting those probes. This is the key concept behind our work: different experiments are sensitive to different physical phenomena, therefore, it is crucial to correctly combine as many independent cosmological probes as possible in order to explore the whole parameter space and avoid any biased determination of the underlying cosmology.
\begin{figure}
    \centering
    \includegraphics[width=\columnwidth]{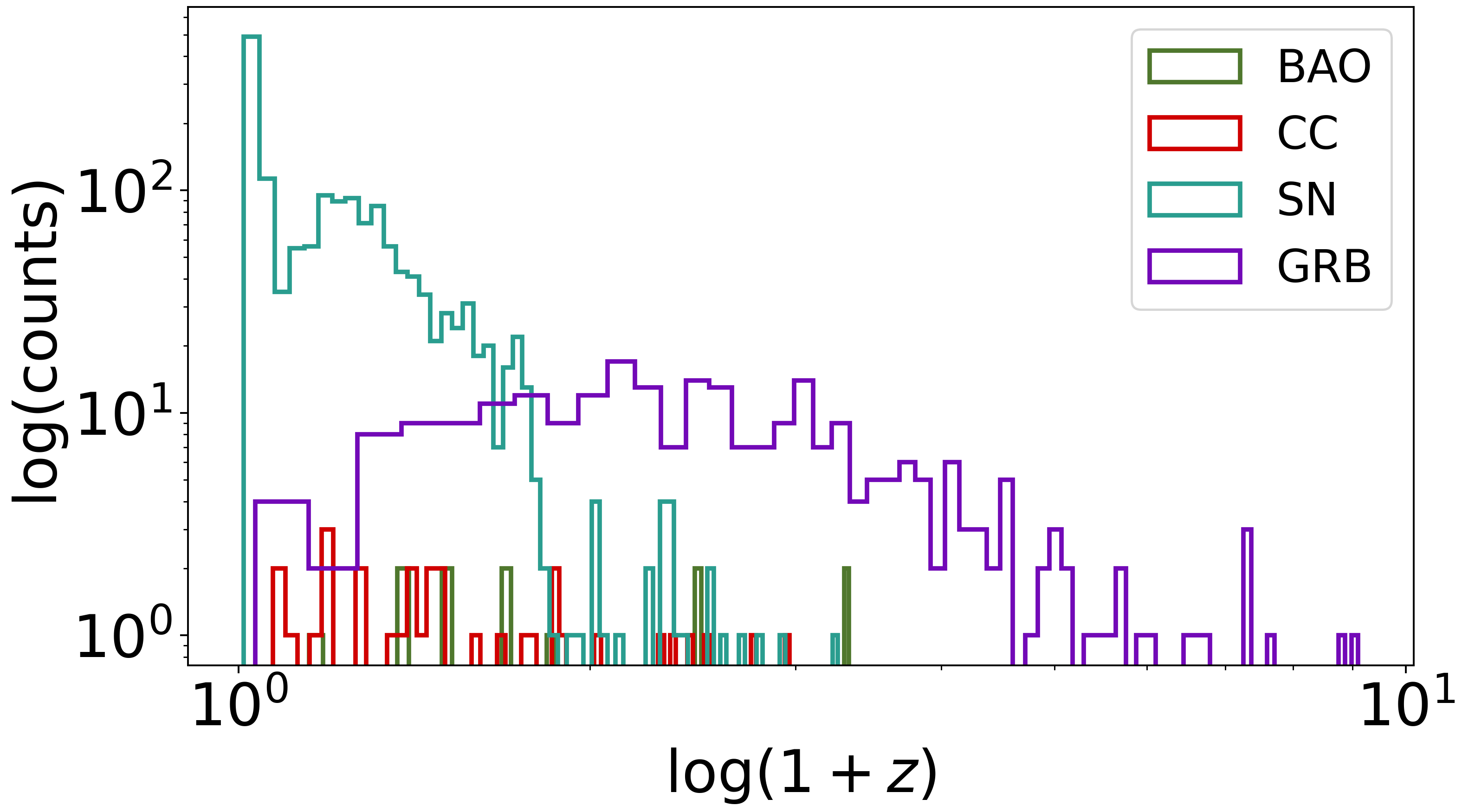}
    \caption{Redshift distribution of data collected in this work.}
    \label{fig:dist_data}
\end{figure}

The paper is organised as follows: in Section \ref{sec:data,model,methods} we briefly describe the main features of each probe and model, as well as the Bayesian framework within which the analysis was carried out.
Section \ref{sec:res} presents the results of the cosmological inference, with 
special attention to the impact and reliability of the probe--by--probe strategy.
Then, in Section \ref{sec:syn_com} an original study of synergies and complementarities between probes is presented, while Section \ref{sec:conclusions} summarizes the main outcomes of our work.

\section{Data, Models \& Methods}
\label{sec:data,model,methods}
\subsection{Late--Universe data}
\label{sec:data}

Combining ``standard rulers, candles and clocks'' is an old recipe \citep{Heavens14} to extract the cosmological signal using exclusively low--redshift data with a model--independent approach (see also \citealt{moresco16b} and \citealt{benitsy20}). Aiming to extend this pioneering work, we select the most updated and complete late--Universe datasets available today.
More specifically, this work is based on the analysis of two primary probes, such as Type Ia SuperNovae and Baryon Acoustic Oscillations, which are combined with two new emerging methodologies, namely Cosmic Chronometers and Gamma--Ray Bursts, that have recently proven their strength and reliability \citep{morescorev22}.

As shown in Figure \ref{fig:dist_data}, our dataset allows us to span a wide redshift range $0<z<2$ with most of the probes, where Gamma--Ray Bursts observations extend it up to $z\sim10$. Hence, we extract the cosmological signal from a time frame currently not deeply explored in the literature.

\subsubsection{Type Ia SuperNovae}
Type Ia SuperNovae (SNe) are one of the most powerful and suggestive astrophysical events observed in the cosmos.
The most recent and complete SNe collection is the so-called Pantheon+ sample \citep{shoes_pantheon+}, in which 1701 observations ($0.001<z<2.26$) of 1550 SNe are collected.
\\
Following \citet{kessler_scolnic17}, the brightness of these objects is standardized through the definition of the distance modulus $\mu$ as:
\begin{equation}
\label{eq:mB}
	\mu = m_B^{\text{corr}} - M
\end{equation}
where $m_B^{\text{corr}}$ is the apparent magnitude taking into account several correction terms, such as selection bias, dust extinction, light--curve colour and stretch parameter, while the nuisance parameter $M$ is the absolute magnitude of a fiducial SNe.
Since $\mu$ is related to the luminosity distance $D_L$ through the following equation:
\begin{equation}
\label{eq:mu_dL}
	\mu(z) = 5\ \text{log}_{10}\bigg( \frac{D_L(z)}{1 Mpc}\bigg) + 25
\end{equation}
by measuring the apparent magnitudes $m_B$ of an SNe sample it is possible to constrain the cosmological parameters thanks to the generic expression:
\begin{equation}
	D_L(z) =\frac{(1+z)}{\sqrt{|\Omega_{k}|}} \ \ S_k\bigg( \sqrt{|\Omega_{k}|} \int_0^z \frac{c\ \text{d}z'}{H(z')}\bigg)
	\label{dL_esplicita}
\end{equation}
where $\Omega_k$ is the curvature parameter, $H(z)$ is the Hubble parameter and the function $S_k$ is defined as:
\begin{equation}
	S_k(\chi) = 
	\begin{cases}
		\text{sin}(\chi) \ \ \ \ \ \text{for} \ \ \ \Omega_k<0 \ \rightarrow{} \ \text{closed Universe},\\
		\ \ \ \chi \ \ \ \ \ \ \ \ \ \ \text{for} \ \ \ \Omega_k=0 \ \ \  \rightarrow{} \ \text{flat Universe},\\
		\text{sinh}(\chi) \ \ \ \text{for} \ \ \ \Omega_k>0 \ \rightarrow{} \ \text{open Universe}.\\
	\end{cases}
\end{equation}
Due to the intrinsic degeneracy between $H_0$ and the nuisance parameter $M$, SNe are not able to constrain the expansion rate of the Universe. At the same time, SNe have been widely used in the past three decades to precisely infer the energy budget of the Universe.

Following the same procedure reported in \citet{shoes_pantheon+}, we do not consider the nearby Hubble Diagram ($z<0.01$) in order to avoid any systematics due to unmodeled peculiar velocities. Therefore, our final cosmological sample includes 1590 observations with their associated covariance matrix.

\subsubsection{Baryon Acoustic Oscillations}
\label{sec:dataBAO}
Density fluctuations in the primordial photo--baryonic fluid affect the formation of large--scale structures in the more recent stages of the Universe. 
In fact, after the photo--baryonic decoupling ($z\sim1100$), the baryons at the boundary of the gravitational potential wells preferably cluster at a distance from the centre of the CDM haloes fixed by the length of the radius $r_d$ of the sound horizon calculated at the time of decoupling.
As a consequence of this phenomenon, in the late--time matter distribution it is most probably to find two galaxies separated by a distance $r_d$, and this particular feature, known as Baryon Acoustic Oscillations (BAOs), represents an incredible and versatile way to infer cosmological parameters. 
Considering $r_d$ as a standard ruler, it is possible to measure different types of observables: the Hubble distance $D_H(z)\equiv c/H(z)$, the transverse comoving distance $D_M(z)\equiv D_L(z)/(1+z)$ or the volume--averaged distance $D_V(z)\equiv[z D_H(z) D_M^2(z)]^{1/3}$, depending on the direction (radial and/or transverse) along which the BAOs signal is observed.

In this framework, we select a sample ($0<z<2.4$) of ``BAO--only'' measurements -- collected in \citet{Alam21} -- carried out from different surveys by \citet{Ross15}, \citet{Alam17}, \citet{gil20}, \citet{dumas21}, \citet{deMattia20} and \citet{eBOSSDR16}. Here, ``BAO--only'' means that the measurements are uncalibrated, i.e., the value of $r_d$ is assumed as a constant parameter to be inferred and is not calibrated through early--time Universe observations.
The observables $D_H$, $D_M$, and $D_V$ are indeed scaled by a factor $r_d^{-1}$ which, as explained by \citet{brieden23}, is assumed to be a constant and isotropic length.

Hence, like SNe, BAOs are a very useful tool for measuring the energy budget of the Universe whilst they are insensitive to the Hubble constant $H_0$.

\subsubsection{Cosmic Chronometers}
The method of Cosmic Chronometers (CCs), introduced for the first time in \citet{jim_loeb}, measures the Hubble parameter as:
\begin{equation}
\label{Hz_CC}
    H(z)=-\frac{1}{1+z}\frac{\text{d}z}{\text{d}t} \ .
\end{equation}
While the redshift $z$ is a directly observable quantity, the variation of the look--back time d$t$ is determined by analysing the spectra or photometry of a particular population of massive and passive galaxies whose majority of stars were formed in the early stages of their evolution.

The reason why this method has been increasingly used in cosmological analyses relies on its independence from cosmology. 
Assuming only the validity of the General Relativity, the Cosmological Principle, and the Weyl postulate, with CCs measurements we are able to directly track the history of the cosmic expansion $H(z)$. Thus, CCs are consolidating their role in the modern cosmological framework thanks to the relatively simple needs and the minimal cosmological assumptions they use.

As summarized in the review by \citet{morescorev22}, there are several methods to derive d$t$ measurements from the CCs observations: the full--spectrum fitting \citep[see, e.g.,][]{jiao2023,tomasetti2023}, the Lick indices analysis \citep[see, e.g.,][]{borghi}, and the calibration of specific spectroscopic features \citep[see, e.g.,][]{moresco2012,moresco2015,moresco16a}. More recently, it has been demonstrated the possibility of using photometric surveys to retrieve accurate $H(z)$ measurements by selecting CCs with a Machine Learning approach \citep{jimenez2023}.

Until now, this method has provided 32 measurements of the Hubble parameter over the redshift range $0.07\leq z \leq 1.965$ \citep{morescorev22}.

\subsubsection{Gamma--Ray Bursts}

Despite their great success in measuring cosmological parameters, the SNe observations do not go beyond $z\sim2$. 
Thanks to the observational efforts of the last decades, a new interesting kind of distance indicator emerged allowing us to investigate stages of the Universe out of the reach of the standard cosmological probes. 
This is the case of long Gamma--Ray Bursts (GRBs), the most energetic explosive events observed in the cosmos, produced by the core--collapse of peculiar massive stars (see, e.g., \citealt{piran05}, \citealt{kumar14} and \citealt{levan16}). The combination of their origin, huge brightness and redshift distribution extending up to more than $z\sim9$ makes these phenomena very powerful probes for cosmology. GRBs are characterized by a prompt emission, lasting typically from a few seconds up to a few minutes, during which most of the energy is radiated in X/$\gamma$--rays, and a multi--wavelength afterglow emission spanning from $\gamma$--rays to radio and fading on time scales ranging from several hours to several days.

Certainly, GRBs are not standard candles. Also, the emission mechanisms at work, especially during the prompt phase, are not yet fully understood and only some aspects of the progenitors models are known. But, the discovery and deep investigation of empirical correlations between radiated energy (or peak luminosity) and spectral (or temporal) properties is consolidating the role of these events in modern observational cosmology (see, e.g., \citealt{amati13}, \citealt{dainotti18}, and for a recent review \citealt{morescorev22}). In this context, the most investigated correlation for GRBs physics and cosmology is by far the ``Amati relation'' \citep{amati02,amati06,amati08,amati13} between the photon energy at which the time--integrated $\nu$F$_\nu$ spectrum of the X/$\gamma$--rays prompt emission peaks, i.e., the cosmological rest--frame peak energy $E_{\text{p,i}} = E_{\text{p}} (1+z)$, and the isotropic--equivalent radiated energy $E_{\text{iso}}$. The $E_{\text{p,i}}$ -- $E_{\text{iso}}$ (``Amati'') correlation takes the form:
\begin{equation}
	\label{eq:rel_amati}
	\text{log}\bigg[\frac{E_{\text{p,i}}}{\text{keV}}\bigg] = b + a \ \text{log}\bigg[\frac{E_{\text{iso}}}{10^{52} \text{ erg}}\bigg]
\end{equation}
where $a$ and $b$ are constants to be inferred.
While $E_{\text{p,i}}$ is directly extracted from the measured prompt spectrum, the quantity $E_{\text{iso}}$ is related to the luminosity distance through the following relation:
\begin{equation}
\label{eq:Eiso_dL}
    \begin{split}
    E_{\text{iso}}\ = & \ \ 4 \pi D_L^2(z) (1+z)^{-1} \int^{10^4/(1+z)}_{1/(1+z)} E\ N(E)\ \text{d}E \ = \\
    \ = & \ \ 4 \pi D_L^2(z) (1+z)^{-1} S_{\text{bolo}}
    \end{split}     
\end{equation}
where $N(E)$ is the Band function \citep{band93} and $S_{\text{bolo}}$ is the measured bolometric fluence.

Therefore, such a relation represents a remarkable tool to infer cosmological parameters since it correlates two observable quantities ($E_\text{p}$ and $S_\text{bolo}$), one of which ($S_\text{bolo}$) depends on them according to Equation \ref{eq:Eiso_dL}.
Although it is very highly significant, the Amati relation (Equation \ref{eq:rel_amati}) is characterized by a scatter of the data around the best--fit power--law which significantly exceeds the level expected from Poissonian fluctuations and underestimated systematics in the measurement of the two quantities. Thus, an additional parameter $\sigma_{\text{int}}$ has to be considered to take into account the intrinsic scatter of the correlation. Moreover, as already stated for BAOs and SNe, the intrinsic degeneracy between $H_0$ and the nuisance parameter $b$ does not allow us to infer the Hubble constant value from the ``Amati'' relation. 

\begin{figure}
    \centering
    \includegraphics[width=\columnwidth]{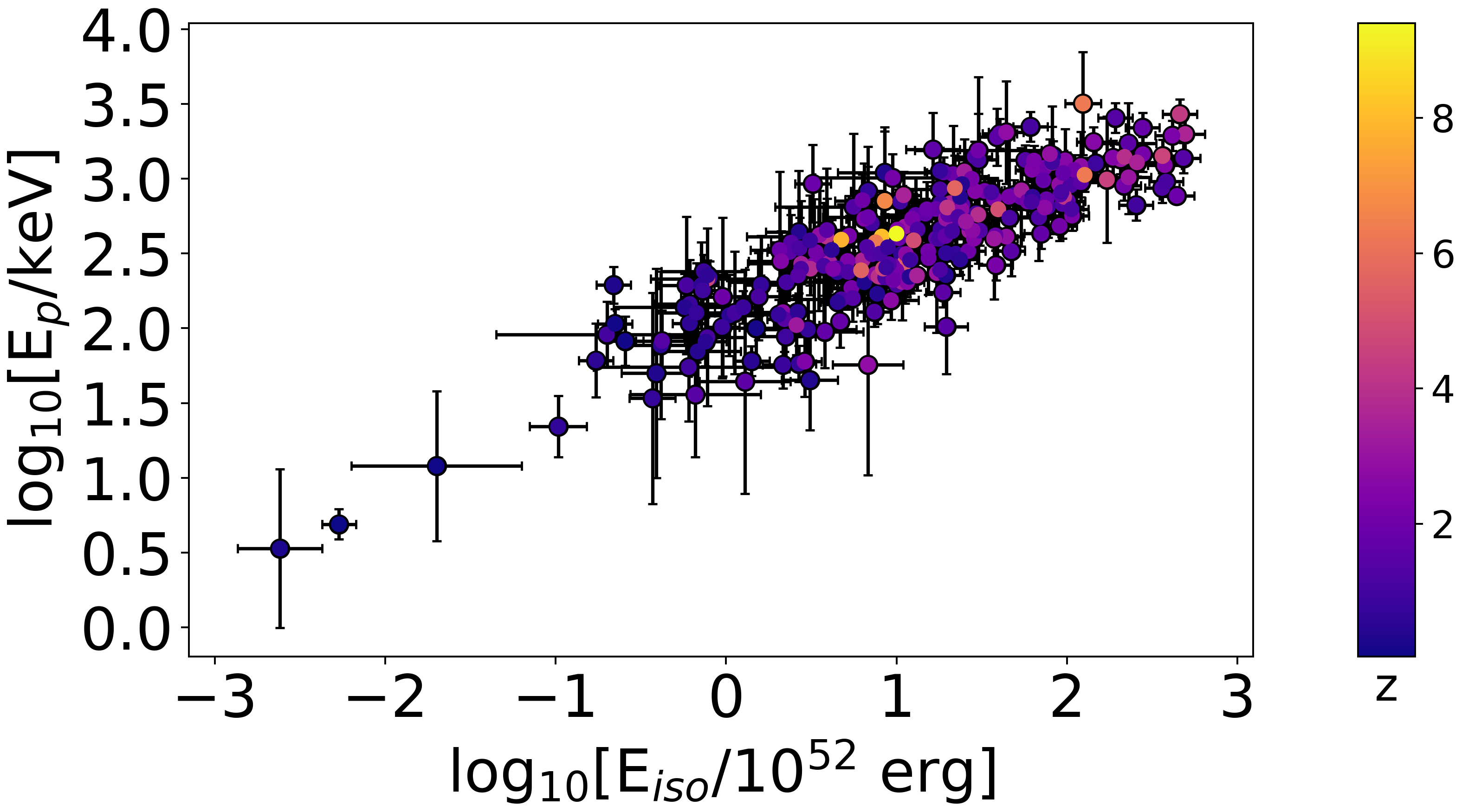}
    \caption{Gamma--Ray Bursts data taken from Amati et al. (in prep.).}
    \label{fig:GRB_data}
\end{figure}

In our multi--probes cosmological analysis, we introduce GRBs through the Amati relation based on the updated $E_{\text{p,i}}$ -- $E_{\text{iso}}$ dataset by Amati et al. (in prep.), which includes 264 GRBs observations up to redshift values $z\sim9$ and is  represented in Figure \ref{fig:GRB_data}.

The values of $E_{\text{p,i}}$ and $E_{\text{iso}}$ of these GRBs are based on measurements of fluence and spectral parameters by the main GRBs detectors operated since the first measurements on GRBs redshifts in 1997 up to end of 2022: CGRO/BATSE, BeppoSAX/GRBM, HETE--2, Swift/BAT, Fermi/GBM and Konus--WIND. The sample was built by updating and integrating the early compilation by \citet{amati08} with the measurements reported in more recent systematic spectral analysis and catalogues \citep{amati09,gruber11,atteia17,tsvetkova17,dirirsa19,katsukura20,minaev20,tsvetkova21} and, for the latest GRBs, in GCN Circulars\footnote{\url{https://gcn.gsfc.nasa.gov/gcn3\_archive.html}}. In order to get the most robust possible sample, selection criteria were applied to the considered datasets, including the total duration of the event, the energy band of the detector, the signal--to--noise ratio of the fluence measurement, and the integration time of the GRBs spectrum with respect to the total burst duration. More details will be provided in Amati et al. (in prep.).

\subsection{Cosmological models}
\label{sec:models}
To provide a detailed study on the extensions of the standard cosmological $\Lambda$CDM model, we focus our attention on a specific family of cosmological models that, in the CDM framework, investigate the null curvature hypothesis ($\Omega_k\equiv0$) and the assumption of Dark Energy as a Cosmological Constant $\Lambda$ ($w\equiv-1$).
In the more general case, the DE EoS is described by the CPL model developed by
\citet{chev} and \citet{linder} in the early 2000s. 
This is a two--parameters model describing the variation with redshift $z$ of the parameter $w$ as:
\begin{equation}
\label{w_cpl}
	w(z) = w_0 + w_a\bigg(\frac{z}{1+z}\bigg) \ .
\end{equation}
Without any a priori assumption on the curvature of the Universe, the evolution of the cosmic expansion is thus described through the following equation:
\begin{equation}
	H(z) = H_0 \sqrt{\Omega_{m} (1+z)^3 + \Omega_k (1+z)^2 +  \Omega_{\Lambda}f(z)}
\label{Hz_ow0waCDM}
\end{equation}
where $f(z)\equiv(1+z)^{3(1+w_0+w_a)}e^{-3w_a\frac{z}{1+z}}$ and $\Omega_k\equiv1-\Omega_m-\Omega_{\Lambda}$. 
Here, even if not indicated, all the components to the energy budget $\Omega_i$ are referred to their value at $z=0$, e.g., $\Omega_m\equiv\Omega_{m,0}$.

Therefore, three different DE EoS are derived by simply fixing particular parameters in Equation \ref{w_cpl}, namely:
\begin{itemize}
    \item $w_0w_a$CDM corresponds to all free parameters;
    \item $w$CDM corresponds to  $w_a\equiv0$ ;
    \item $\Lambda$CDM corresponds to $w_0\equiv-1$ and $w_a\equiv0$;
\end{itemize}
and, for each of them, the null (or flat) curvature case is investigated as well by imposing $\Omega_{\Lambda}\equiv1-\Omega_m$ in Equation \ref{Hz_ow0waCDM}.

\subsection{Bayesian framework}
\label{sec:meth}
To constrain cosmological parameters we base our strategy on a statistical approach known as Markov Chain Monte Carlo (MCMC).
In this framework, once the likelihood functions are defined and the posterior distributions are constructed through the priors definition (reported in Table \ref{tab:prior_std}), a random walk into the parameter space determines the best--fit values of the parameters. 

In the present paper, we apply the  widely used \texttt{emcee} software, i.e., a pure--Python implementation of the Goodman $\&$ Weare’s MCMC sampler \citep{emcee}.
We verified the convergence of our chains considering the Gelman--Rubin criterion with $R<0.01$. 
\begin{table*}
\caption{Priors on the cosmological parameters explored with the MCMC method in the analyses involving BAOs, CCs, SNe and GRBs data. The symbol $\mathcal{U}$ indicates a uniform prior between the indicated extremes.}
\label{tab:prior_std}
\renewcommand{\arraystretch}{1.5} 
     \begin{center}
     \begin{tabular}{ccccc|ccccc}
     \hline
     \multicolumn{5}{c|}{Cosmological Parameters} & \multicolumn{5}{c}{Nuisance Parameters} \\ [0.3ex]
     \hline
	$H_0$ & $\Omega_m$ & $\Omega_{\Lambda}$  & $w_0$ & $w_a$  & $r_d$ & $M$ & $a$ & $b$  & $\sigma_{\text{int}}$ \\ [0.3ex] 
    [km s$^{-1}$ Mpc$^{-1}$] & -- & --  & -- & --  & [Mpc] & -- & -- & --  & -- \\ [0.3ex]
	 \hline
     \hline
     $\mathcal{U}$\big(0, 100\big) & $\mathcal{U}$\big(0, 1\big) & $\mathcal{U}$\big(0, 1\big) & $\mathcal{U}$\big($-5$, $-0.3$\big) & $\mathcal{U}$\big($-5$, 5\big) & $\mathcal{U}$\big(50, 250\big) & $\mathcal{U}$\big(-25, -15\big) & $\mathcal{U}$\big(0, 3\big) & $\mathcal{U}$\big(0, 5\big) & $\mathcal{U}$\big(0, 1\big) \\ [0.3ex] 
     \hline
     \end{tabular}
     \end{center}
\end{table*}
\begin{figure*}
    \centering
    \begin{subfigure}[t]{0.39\textwidth}
    \includegraphics[width=\textwidth]{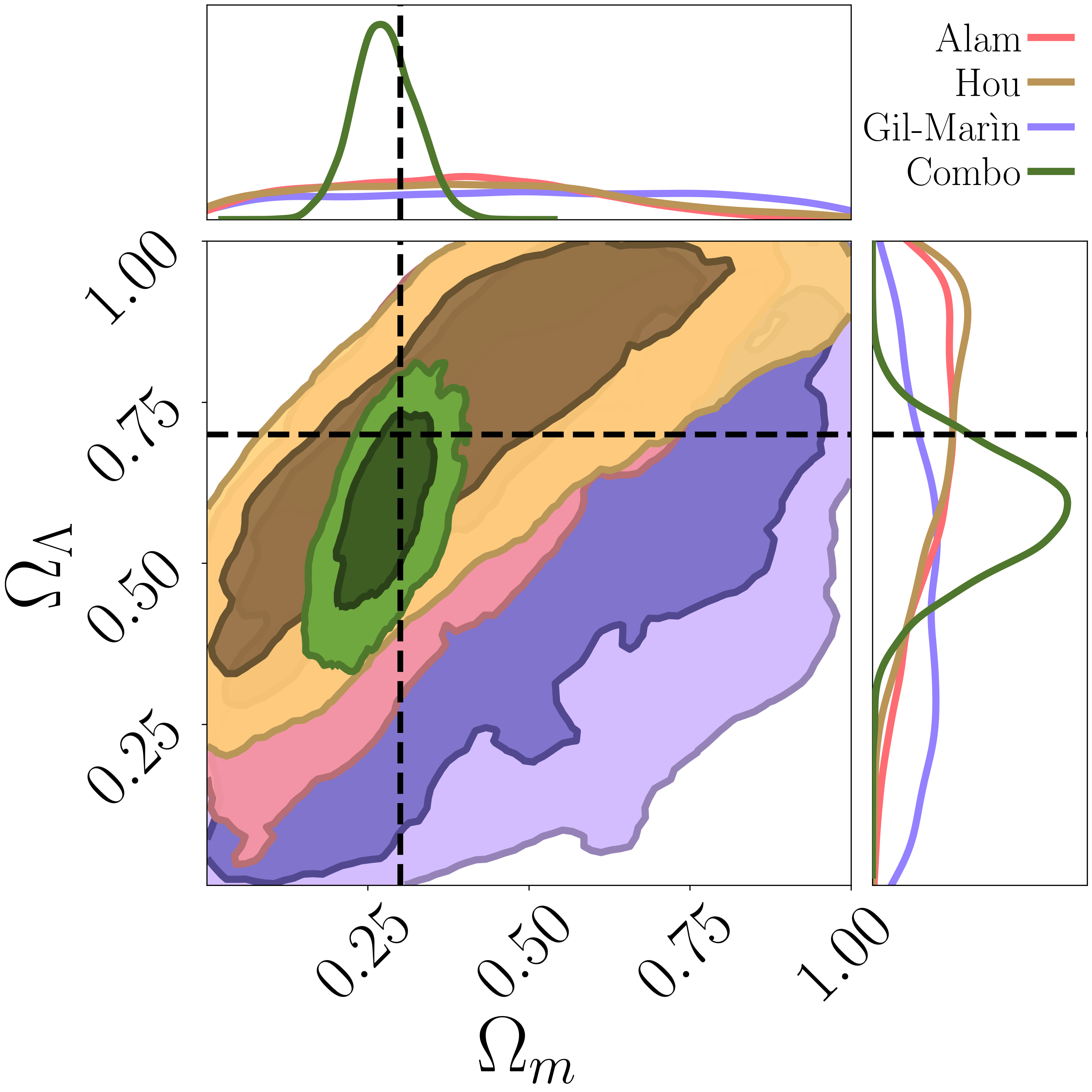}
    \caption{$\Lambda$CDM}
    \label{fig:BAO_resFLCDM}
    \end{subfigure}
    \begin{subfigure}[t]{0.395\textwidth}
    \includegraphics[width=\textwidth]{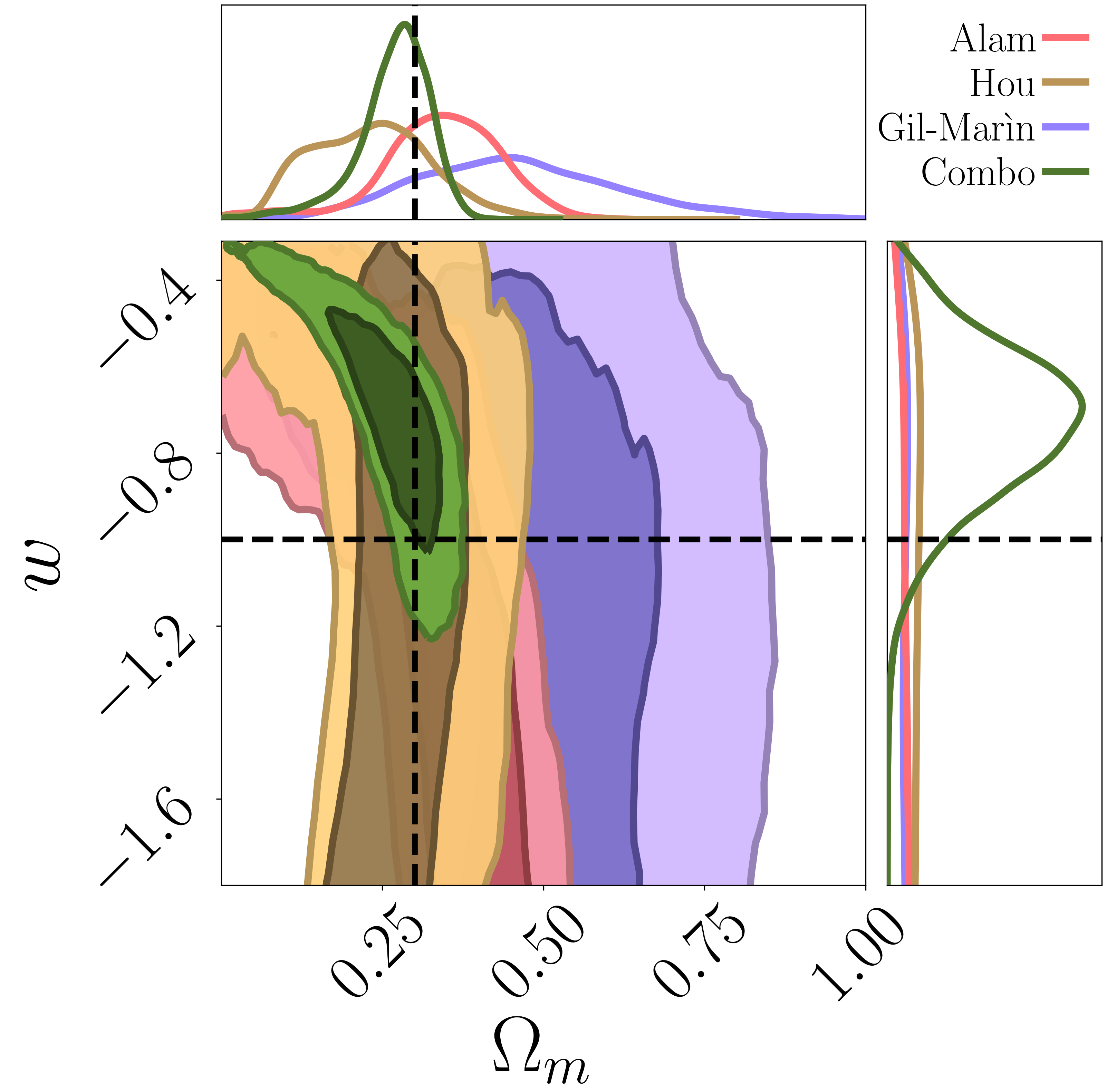}
    \caption{flat $w$CDM}
    \label{fig:BAO_resFwCDM}
    \end{subfigure}
    \caption{Contour plots (at $68\%$ and $95\%$ C.L.) and one--dimensional marginalized distribution inferred from \citet{Alam17}, \citet{eBOSSDR16}, \citet{gil20} and the combination of all BAOs data collected in this work. The left panel shows the $\Omega_m-\Omega_{\Lambda}$ plane ($\Lambda$CDM), while the right panel shows the $\Omega_m-w$ plane (flat $w$CDM). Dashed black lines represent the reference value of each parameter, $\Omega_m=0.3$, $\Omega_{\Lambda}=0.7$ and $w=-1.0$.}
    \label{fig:BAO_res}
\end{figure*}

We consider more conservative priors on nuisance parameters w.r.t. those from the literature. In particular, the prior on $M$ is roughly 5 times wider than the one imposed by \citet{shoes_pantheon+}.
Moreover, to avoid biases, the starting points of the MCMC have been selected with a random extraction within the prior intervals.

The best--fit value $\hat{\theta}$ and the 68\% (95\%) Confidence Level (C.L.) of each parameter are extracted from the sampled distribution by marginalizing over the other ones and taking the $50^{\text{th}}$ percentile and $16^{\text{th}}$ ($2.5^{\text{th}}$) and $84^{\text{th}}$ ($97.5^{\text{th}}$) percentiles, respectively.

\section{Cosmological constraints}
\label{sec:res}
\subsection{Single probes}
\label{sec:analysis_sing}
As a first step, we test the relative strength and constraining power of each separate cosmological probe.
\subsubsection{Baryon Acoustic Oscillations}
\label{sec:BAO}
As already discussed, our BAOs sample is composed of a diverse set of observations coming from a wide variety of spectroscopic surveys and cosmological quantities, namely $D_H(z)/r_d$,  $D_M(z)/r_d$ and  $D_V(z)/r_d$.

We first analysed the individual BAOs datasets and compared the derived constraints, to verify the consistency of the results. \citet{Alam21} verified that the correlation between data is negligible and, moreover, provided a collection of covariance matrices for those sub--samples whose systematics were computed, namely, \citet{Alam17}, \citet{gil20} and \citet{eBOSSDR16}.
To demonstrate how the combination of different BAOs probes allows us to significantly improve the cosmological measurements, in Figure \ref{fig:BAO_res} we show the constraints obtained from the main (separate) BAOs data, revealing how the synergies between various observables could be fundamental to achieving better precision in the inference process.
Then, to analyse the full BAOs datasets we combine the likelihood functions of the different measurements as:
\begin{equation} \label{eq:like_BAO}
\begin{split}
	\text{ln}\ &\mathcal{L}_{\text{BAO}}\ =   \ \text{ln}\ \mathcal{L}_{\text{Ross}}\ +\  \text{ln}\ \mathcal{L}_{\text{Alam}}\ +\ \text{ln}\ \mathcal{L}_{\text{Gil--Mar{\'\i}n}}\ +  \\
    & +\ \text{ln}\ \mathcal{L}_{\text{du Mas de Bourboux}} \ + \text{ln}\ \mathcal{L}_{\text{de Mattia}}\ + \text{ln}\ \mathcal{L}_{\text{Hou}}
\end{split}
\end{equation}
where the exact form of the likelihood function depends on whether or not the sub--sample is provided with a covariance matrix, that is:
\begin{equation}
	\label{eq:like_gauss}
	\text{ln}\ \mathcal{L}\ \propto \ -\frac{1}{2} \sum^N_{i=1} \bigg[\frac{x_i - y_i}{\sigma_i}\bigg]^2
\end{equation}
or:
\begin{equation}
\label{likelihood_cov}
	\text{ln}\ \mathcal{L}\ \propto \ -\frac{1}{2}\sum^N_{i=1}\sum^N_{j=1}\bigg[ \big(x_i - y_i\big)\ Cov^{-1}_{ij}\ \big(x_j - y_j\big)\bigg]
\end{equation}
where the sum runs on the $N$ measurements in the dataset. 
\\
$y_i\equiv y(\theta, z_i)$ represents the theoretical value -- depending on the cosmological parameters $\theta$ -- of the observable $x_i\equiv x(z_i)$. Finally, $\sigma_i$ is the Gaussian--assumed error of the $i-$th measurement and $Cov_{ij}$ is the matrix describing the variation of the $i-$th measurement with respect to the $j-$th one.

In Table \ref{tab:res_conclusioni} we report the main cosmological constraints from BAOs data, among which we highlight the significance of the inferred values of $\Omega_m$.
On the other hand, this BAOs collection does not precisely constrain $\Omega_{\Lambda}$ and the DE EoS parameters $w_0$ and $w_a$, because of the strong degeneracy between parameters that quickly grows as the model dimension increases.
Our results fully agree with those carried out by \citet{Alam21}\footnote{Publicly available here: \url{https://svn.sdss.org/public/data/eboss/DR16cosmo/tags/v1_0_1/}}.

\subsubsection{Type Ia SuperNovae}
\label{sec:SNe}
As explained above, \citet{shoes_pantheon+} applied a cut to the full Pantheon+ sample and to the associated covariance matrix. Following their approach, our analysis is restricted to 1590 observations from which we construct the likelihood function as:
\begin{equation}
\label{eq:like_SN}
	\text{ln}\ \mathcal{L}_{\text{SN}}\ \propto \ -\frac{1}{2}\sum^{1590}_{i=1}\sum^{1590}_{j=1}\bigg[ (m^{\text{corr}}_{B} - \mu_{th})_i\ \cdot Cov^{-1}_{ij}\ \cdot(m^{\text{corr}}_{B} - \mu_{th})_j\bigg]
\end{equation}
where $\mu_{th}=5\ \text{log}_{10}D_L(\theta,z) + M$, with $\theta$ and $M$ respectively representing the cosmological and nuisance parameters.
\\
Table \ref{tab:res_conclusioni} shows the main results of the SNe analysis. Through the most updated SNe sample, we derive precise constraints on both the density parameters $\Omega_m$ and $\Omega_{\Lambda}$ but also on the DE EoS parameter $w_0$. Obviously, such precision shows a decreasing trend as the complexity of the model increases and the cosmological parameters become more degenerate with each other.
As already mentioned and widely demonstrated in the literature, SNe do not constrain the Hubble constant as this parameter is strongly degenerate with $M$. A more detailed analysis of this feature is given in Section \ref{sec:nuisance_vs_H0}.

The agreement between our results and those obtained by the original authors\footnote{Publicly available here: \url{https://github.com/PantheonPlusSH0ES/DataRelease/tree/main/Pantheon\%2B_Data/5_COSMOLOGY/chains}} further demonstrates the reliability of our implementation of the Bayesian approach. Moreover, we extend their analysis also to the $w$CDM and $w_0w_a$CDM models, finding no significant deviations from the standard cosmological model.
\begin{table*}
	\begin{center}
	\caption{Main cosmological results (best--fit and 68$\%$ C.L. values) from the analyses of BAOs, SNe, CCs, and GRBs data. Note that $r_d$ and $H_0$ values are respectively in units of Mpc and km s$^{-1}$ Mpc$^{-1}$. Empty values correspond to parameters not constrained by the corresponding cosmological model. The values of $\Omega_k (\equiv 1 - \Omega_m - \Omega_{\Lambda})$ are directly extracted by combining the marginalized distributions of $\Omega_m$ and $\Omega_{\Lambda}$.}
		\label{tab:res_conclusioni}
		    \renewcommand{\arraystretch}{1.5}
			\begin{tabular}{c|cccccc}
			\hline
				 & flat $\Lambda$CDM & $\Lambda$CDM & flat $w$CDM & $w$CDM & flat $w_0w_a$CDM & $w_0w_a$CDM  \\ 
			    \hline
				\hline	
				\multicolumn{1}{c|}{} & \multicolumn{6}{c}{\textbf{BAO}} \\
				\hline
                 $\Omega_m$ & $0.316^{+0.030}_{-0.027}$ & $0.274^{+0.051}_{-0.046}$ & $0.278^{+0.044}_{-0.051}$ & $0.244^{+0.058}_{-0.058}$ & $0.299^{+0.057}_{-0.104}$ & $0.245^{+0.077}_{-0.094}$ \\ [0.3ex] 
                 \hline
                 $\Omega_{\Lambda}$ & -- & $0.581^{+0.094}_{-0.098}$ & -- & $0.632^{+0.135}_{-0.132}$ & -- & $0.593^{+0.147}_{-0.138}$ \\ [0.3ex]  
                 \hline
                 $\Omega_{k}$ & -- & $0.142^{+0.134}_{-0.130}$ & -- & $0.122^{+0.163}_{-0.156}$ & -- & $0.159^{+0.168}_{-0.161}$ \\ [0.3ex] 
                 \hline
                $w_0$ & -- & -- & $-0.73^{+0.16}_{-0.19}$ & $-0.77^{+0.20}_{-0.27}$ & $-0.69^{+0.23}_{-0.27}$ & $-0.76^{+0.26}_{-0.55}$ \\ [0.3ex] 
                \hline
                $r_d$ & $160.4^{+62.3}_{-47.7}$ & $163.0^{+59.6}_{-45.0}$ & $157.6^{+59.5}_{-42.6}$ & $152.2^{+67.0}_{-40.2}$ & $159.4^{+57.7}_{-47.9}$ & $153.5^{+59.0}_{-41.9}$ \\ [0.3ex] 
                
				\hline
				\hline
				\multicolumn{1}{c|}{} & \multicolumn{6}{c}{\textbf{SN}} \\
				\hline
				$\Omega_m$ & $0.331^{+0.018}_{-0.017}$ & $0.300^{+0.053}_{-0.057}$ & $0.296^{+0.064}_{-0.085}$ & $0.276^{+0.058}_{-0.066}$ & $0.340^{+0.087}_{-0.149}$ & $0.238^{+0.096}_{-0.103}$ \\ [0.3ex]  
                 \hline
                 $\Omega_{\Lambda}$ & -- & $0.616^{+0.080}_{-0.084}$ & -- & $0.484^{+0.329}_{-0.197}$ & -- & $0.364^{+0.201}_{-0.110}$ \\ [0.3ex]  
                 \hline
                 $\Omega_{k}$ & -- & $0.083^{+0.136}_{-0.126}$ & -- & $0.215^{+0.260}_{-0.307}$ & -- & $0.392^{+0.167}_{-0.278}$ \\ [0.3ex] 
                 \hline
                 $w_0$ & -- & -- & $-0.91^{+0.16}_{-0.17}$ & $-1.24^{+0.42}_{-0.72}$ & $-0.92^{+0.15}_{-0.17}$ & $-1.50^{+0.53}_{-0.75}$ \\ [0.3ex]
                 \hline
                $M$ & $-19.2^{+0.4}_{-0.6}$ & $-19.2^{+0.4}_{-0.6}$ & $-19.1^{+0.4}_{-0.7}$ & $-19.3^{+0.5}_{-0.6}$ & $-19.2^{+0.5}_{-0.5}$ & $-19.2^{+0.5}_{-0.6}$ \\ [0.3ex]
                 \hline
				\hline
				\multicolumn{1}{c|}{} & \multicolumn{6}{c}{\textbf{CC}} \\
				\hline
				$H_0$ & $66.4^{+5.3}_{-5.2}$ & $65.6^{+5.6}_{-5.5}$ & $70.7^{+10.3}_{-8.0}$ & $69.7^{+11.2}_{-8.5}$ & $73.1^{+12.4}_{-8.8}$ & $69.9^{+11.2}_{-8.2}$ \\ [0.3ex] 
                 \hline
                $\Omega_m$ & $0.337^{+0.075}_{-0.063}$ & $0.313^{+0.151}_{-0.164}$ & $0.304^{+0.083}_{-0.071}$ & $0.173^{+0.163}_{-0.100}$ & $0.301^{+0.090}_{-0.079}$ & $0.154^{+0.158}_{-0.093}$ \\ [0.3ex] 
                 \hline
                $\Omega_{\Lambda}$ & -- & $0.589^{+0.268}_{-0.303}$ & -- & $0.459^{+0.233}_{-0.239}$ & -- & $0.444^{+0.228}_{-0.240}$ \\ [0.3ex]  
                 \hline
               $\Omega_{k}$ & -- & $0.101^{+0.436}_{-0.399}$ & -- & $0.372^{+0.276}_{-0.354}$ & -- & $0.402^{+0.273}_{-0.322}$ \\ [0.3ex] 
				\hline
				\hline
				\multicolumn{1}{c|}{} & \multicolumn{6}{c}{\textbf{GRB}} \\
				\hline
                $\Omega_m$ & $0.317^{+0.211}_{-0.133}$ & $0.341^{+0.189}_{-0.135}$ & $0.299^{+0.245}_{-0.148}$ & $0.288^{+0.164}_{-0.117}$ & $0.305^{+0.265}_{-0.140}$ & $0.297^{+0.178}_{-0.126}$ \\ [0.3ex] 
                 \hline
                $\Omega_{\Lambda}$ & -- & $0.347^{+0.335}_{-0.244}$ & -- & $0.358^{+0.305}_{-0.241}$ & -- & $0.328^{+0.289}_{-0.226}$ \\ [0.3ex] 
                 \hline
                $\Omega_{k}$ & -- & $0.245^{+0.278}_{-0.270}$ & -- & $0.300^{+0.241}_{-0.231}$ & -- & $0.324^{+0.227}_{-0.230}$ \\ [0.3ex] 
                 \hline
                $w_0$ & -- & -- & $-2.21^{+1.46}_{-1.83}$ & $-2.58^{+1.59}_{-1.70}$ & $-2.68^{+1.54}_{-1.57}$ & $-2.79^{+1.59}_{-1.43}$ \\ [0.3ex]
                \hline
                $a$ & $0.478^{+0.022}_{-0.019}$ & $0.478^{+0.019}_{-0.021}$ & $0.48^{+0.021}_{-0.021}$ & $0.475^{+0.02}_{-0.019}$ & $0.48^{+0.021}_{-0.02}$ & $0.477^{+0.022}_{-0.021}$ \\ [0.3ex]
                 \hline
                 $b$ & $1.95^{+0.20}_{-0.36}$ & $2.0^{+0.17}_{-0.31}$ & $1.94^{+0.19}_{-0.36}$ & $1.98^{+0.18}_{-0.32}$ & $1.92^{+0.19}_{-0.31}$ & $1.98^{+0.18}_{-0.28}$ \\ [0.3ex]
                 \hline
                 $\sigma_{int}$ & $0.213^{+0.011}_{-0.011}$ & $0.213^{+0.011}_{-0.011}$ & $0.214^{+0.012}_{-0.012}$ & $0.213^{+0.011}_{-0.011}$ & $0.214^{+0.011}_{-0.011}$ & $0.214^{+0.012}_{-0.011}$ \\ [0.3ex]
				\hline
				\hline
			\end{tabular}
		\end{center}
\end{table*}
\subsubsection{Cosmic Chronometers}
\label{sec:CC}
As proof of the great effort in the last decade to make this probe increasingly reliable and robust, \citet{moresco20} compute the full covariance matrix for CCs taking into account several potential systematic errors, such as young stellar population, uncertainty in the stellar population synthesis models, and estimate of the stellar metallicity. 
Exploiting the framework built into that work, we write the likelihood function as:
\begin{equation}
\label{eq:like_CC}
	\text{ln}\ \mathcal{L}_{\text{CC}}\ \propto \ -\frac{1}{2}\sum^{32}_{i=1}\sum^{32}_{j=1}\bigg[ (H_{obs} - H_{th})_i\ \cdot Cov^{-1}_{ij}\ \cdot (H_{obs} - H_{th})_j\bigg]
\end{equation}
from which it is possible to infer cosmological parameters considering both statistical and systematic uncertainties.
The analyses of the $\Lambda$CDM models show how this method could provide $H_0$ measurements with a precision up to $\sim9\%$, although far from the one obtained by the current main probes -- such as local Cepheid variables \citep[SH0ES,][]{shoes, shoes_pantheon+} or CMB \citep{planck18}.
Unfortunately, the current number of observations is not enough to keep such good precision when increasing the number of free parameters of the model. Therefore, when used to probe more complicated extensions, this method becomes less sensitive to the physical properties of the Universe, such as the curvature or the DE EoS.
It is worth noting how CCs are more sensitive to the degeneracy between $\Omega_m$ and $\Omega_{\Lambda}$ than to the one between $\Omega_m$ and the DE EoS parameters. In fact, the inferred value of $\Omega_m$ in the flat $w_0w_a$CDM case shows a significantly higher level of precision compared to the one obtained in the $\Lambda$CDM model.
The results reported in Table \ref{tab:res_conclusioni} are in agreement with those coming from the literature (e.g., \citealt{moresco16a, moresco16b} and \citealt{morescorev22}).

\subsubsection{Gamma--Ray Bursts}
\label{sec:GRB}
To extract the cosmological signal from this probe, we exploit the Amati relation based on the most updated and robust dataset, as described in Section \ref{sec:data}. Following the same approach adopted by \citet{amati08} and \citet{amati13}, we construct the likelihood function as:
\begin{equation}
\label{eq:like_GRB}
\begin{split}
\text{ln}\ \mathcal{L}_{\text{GRB}}\  & \propto   \ \frac{1}{2}\sum^{263}_{i=1}\Bigg\{ \text{ln}\bigg[\frac{1+a^2}{2\pi(\sigma^2_{\text{p,i}}+\sigma^2_{\text{iso}}+\sigma^2_{\text{int}})}\bigg]\ + \\ 
\ & - \frac{\Big[\text{log}E_{\text{p,i}} - a \ \text{log}E_{\text{iso}} - b\Big]^2}{\sigma^2_{\text{p,i}}+\sigma^2_{\text{iso}}+\sigma^2_{\text{int}}} \Bigg\}
\end{split}
\end{equation}
where $\sigma_{\text{p,i}}= \frac{\sigma_{E_{\text{p,i}}}}{\text{ln}(10)\cdot E_{\text{p,i}}}$ \big($\sigma_{\text{iso}}  = \frac{\sigma_{E_{\text{iso}}}}{\text{ln}(10)\cdot E_{\text{iso}}}$\big) and  $\sigma_{E_{\text{p,i}}}$ ($\sigma_{E_{\text{iso}}}$) represent the uncertainty on the measurement of $E_{\text{p,i}}$ ($E_{\text{iso}}$).
We refer the reader to \citet{reichart01} for more details about the definition of the likelihood function.
Since the values of $E_{\text{iso}}$ shown in Figure \ref{fig:GRB_data} are calculated from Equation \ref{eq:Eiso_dL} by assuming a fiducial flat $\Lambda$CDM cosmology with $H_0=70$ km s$^{-1}$ Mpc$^{-1}$ and $\Omega_m=0.3$, we impose:
\begin{equation}
    E_{\text{iso}} = E_{\text{iso}}^{\text{fid}} \cdot \Big[\frac{D_L(\theta,z)}{D_L(H_0= 70 \ \text{km s$^{-1}$ Mpc$^{-1}$},\ \Omega_m=0.3,\ z)}\Big]^2
\end{equation}
where $\theta$ are the parameters of the generic model and $E_{\text{iso}}^{\text{fid}}$ are those measurements shown in Figure \ref{fig:GRB_data}.
In this way, we are thus able to avoid any problem of circularity and to constrain the cosmological parameters in a way as reliable as robust.

The results from the analyses of GRBs are reported in the bottom panel of Table \ref{tab:res_conclusioni}. Our work extends the ones from the literature by investigating a wider class of cosmological models, finding results fully consistent with those obtained by \citet{amati08}, \citet{amati13} and \citet{morescorev22}. 
Considering the flat $\Lambda$CDM model we further confirm that also GRBs indicate a value of $\Omega_m$ around $0.3$, even if the level of precision is not comparable with the other probes.
We find, as expected, that for more complicated models the constraining power of this probe decreases.
At the same time, the prospect of probing the cosmos up to a very high redshift ($z\gg1$) makes GRBs one of the most promising emerging probes. Indeed, while on the one side, they allow us to study a phase of the Universe not yet sampled by standard probes, on the other side, these events reveal completely different degeneracy directions compared to other probes, which can be exploited to break degeneracies among parameters and achieve more accurate constraints. 
For example, looking at the orientation of the two--dimensional contours in the $\Omega_m-\Omega_{\Lambda}$ plane ($\Lambda$CDM) reported in Figure \ref{fig:resLCDM}, it should be evident the different orientations of the confidence contours of GRBs and SNe. Although the cosmological observable -- the luminosity distance $D_L(z)$ -- is the same, such a difference in the orientation of the contour is essentially due to the different redshift distributions (see Figure \ref{fig:dist_data}), as well as to the distinct nuisance parameters these two probes adopted to constrain the cosmological ones.

\subsection{Towards the full probes combination}
\label{sec:towards}
\begin{figure*}
    \centering
    \includegraphics[width=\textwidth]{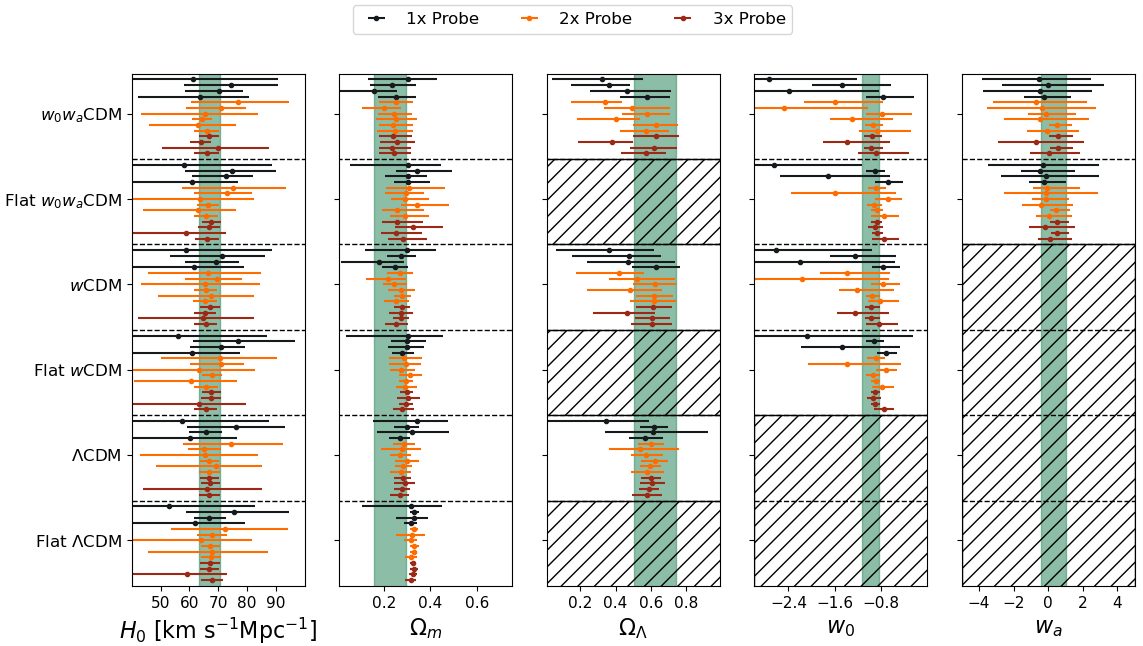}
  \caption{Distribution of the best--fit value (and 68$\%$ C.L.) of cosmological parameters inferred from the progressive combination of probes. Each column shows the constraints obtained for the different cosmological parameters as a function of the cosmological model analysed. Dots with different colours indicate a different number of probes involved in the analysis: individual probes (1x) in dark blue, a couple of probes (2x) in orange, and a triplet of probes (3x) in red. In particular, from top to bottom, each panel (i.e., each model) reports the measurements obtained by GRB, SN, CC, BAO, SN+GRB, CC+GRB, BAO+GRB, CC+SN, BAO+SN, BAO+CC, BAO+CC+SN, CC+SN+GRB, BAO+SN+GRB and BAO+CC+GRB.
  Vertical shaded regions report the value of cosmological parameters inferred from BAO+CC+SN+GRB in the most general $w_0w_a$CDM model (see Table \ref{tab:res_BCSG}). Regions with diagonal line patterns indicate those cosmologies not constraining those specific parameters, e.g., the $\Omega_{\Lambda}$ parameter is not constrained by ``flat'' models.}
    \label{fig:dist_res}
\end{figure*}

After analysing the individual probes separately, we apply a probe--by--probe combination technique to highlight the different constraining powers attainable from the various combinations of cosmological probes.

In Figure \ref{fig:dist_res} the distribution of the best--fit values (and 68$\%$ C.L.) obtained with this approach is shown as a function of the entire set of cosmological parameters.
Two things catch immediately the eye. First, we note that there is a small variance and good agreement between the constraints derived from different combinations of probes.
Then, it is evident how the error bars decrease as the number of probes involved in the combination increases. An exception can be found in the $H_0$ column where the red dots (triplets) corresponding to the BAO+SN+GRB combination show an uncertainty significantly greater than the yellow dots (doublets). However, this behaviour is expected since, as already highlighted, we are considering probes mostly insensitive to the Hubble constant.

This analysis is fundamental to assess the degeneracies affecting each probe (and their combination) and, also, how these features impact cosmological inference and mutate according to the models.
Indeed, the dispersion and the uncertainties of measurements increase as the cosmological model becomes more complicated, as can be seen from the comparison of the flat $\Lambda$CDM and $w_0w_a$CDM distributions of the $\Omega_m$ measurements.

As can be also deduced from Table \ref{tab:res_conclusioni}, we note a marginal trend of our results (at 68$\%$ C.L.) of preferring solutions with hyperbolic geometry ($\Omega_k>0$, $\Omega_{\Lambda}<0.7$) when the curvature of the Universe is left free, as also seen by \citet{moresco16b, shoes_pantheon+, Alam17}. However, at 95$\%$ C.L. the curvature constraints are compatible with a flat geometry.

Furthermore, regarding the $w_0$ distribution, it is worth noticing that increasing the number of combined probes the inferred value tends to $w_0\sim-1$.
Moreover, we note that there are combinations of probes that are less constraining (e.g., CC+GRB) but, also in these cases, the results agree with a Dark Energy component in the form of a Cosmological Constant $\Lambda$.

Finally, looking at the $w_a$ distribution and the large error bars, it should be evident that current cosmological data are not able to accurately constrain this parameter, even if they exclude a significant part of the parameter space, namely, $w_a<-1$ and $w_a>2$.

\subsection{The full combination of probes}
\label{sec:full}
From the probe--by--probe combination analysis, we found a generic consistency between the results obtained from the various methods, so the next step is to simultaneously combine all probes to further increase the precision of the inference.

\begin{figure*}
    \centering
    \begin{subfigure}[t]{0.395\textwidth}
    \includegraphics[width=\textwidth]{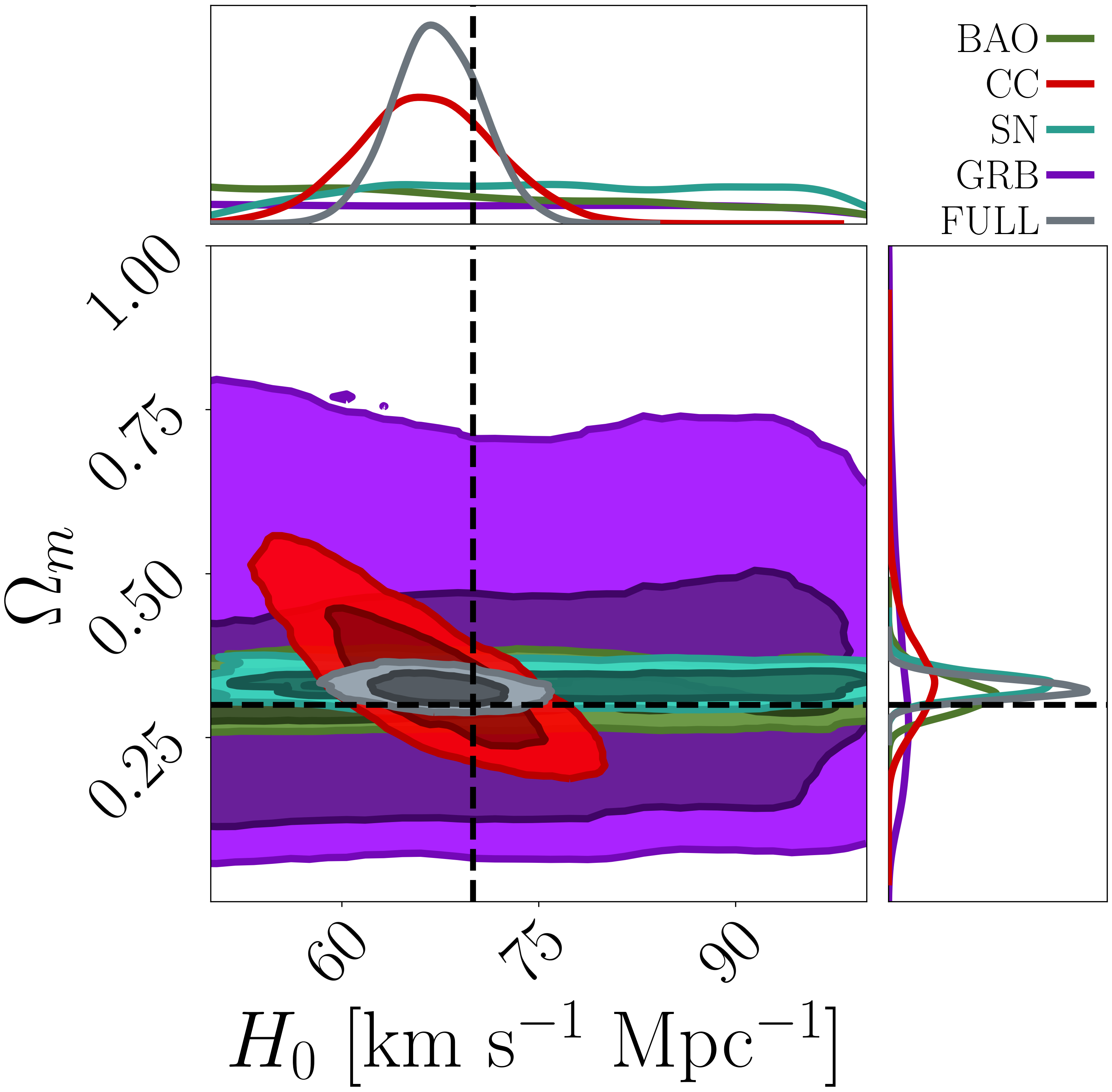}
    \caption{flat $\Lambda$CDM}
    \label{fig:resFLCDM}
    \end{subfigure}
    \begin{subfigure}[t]{0.388\textwidth}
    \includegraphics[width=\textwidth]{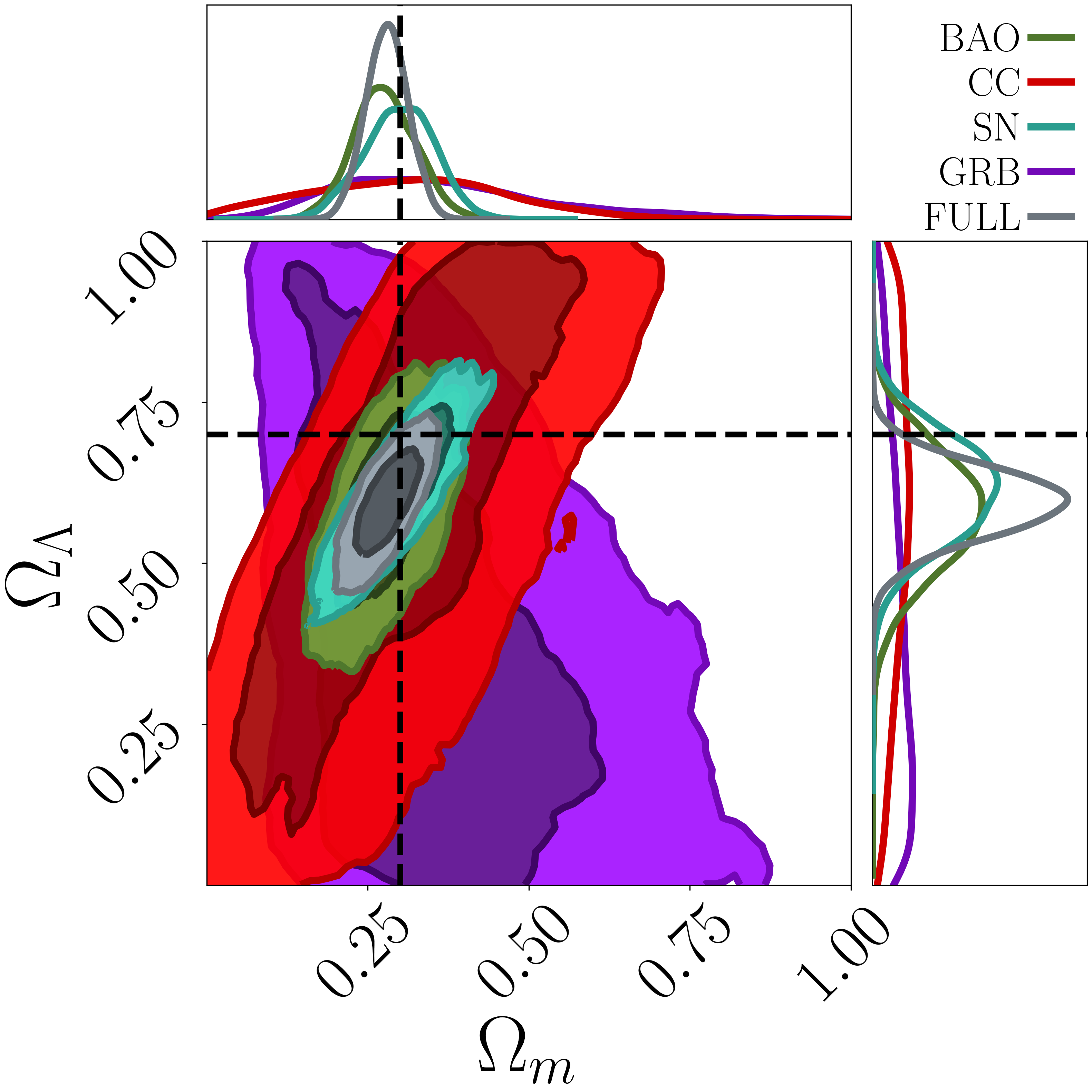}
    \caption{$\Lambda$CDM}
    \label{fig:resLCDM}
    \end{subfigure}
    \begin{subfigure}[t]{0.395\textwidth}
    \includegraphics[width=\textwidth]{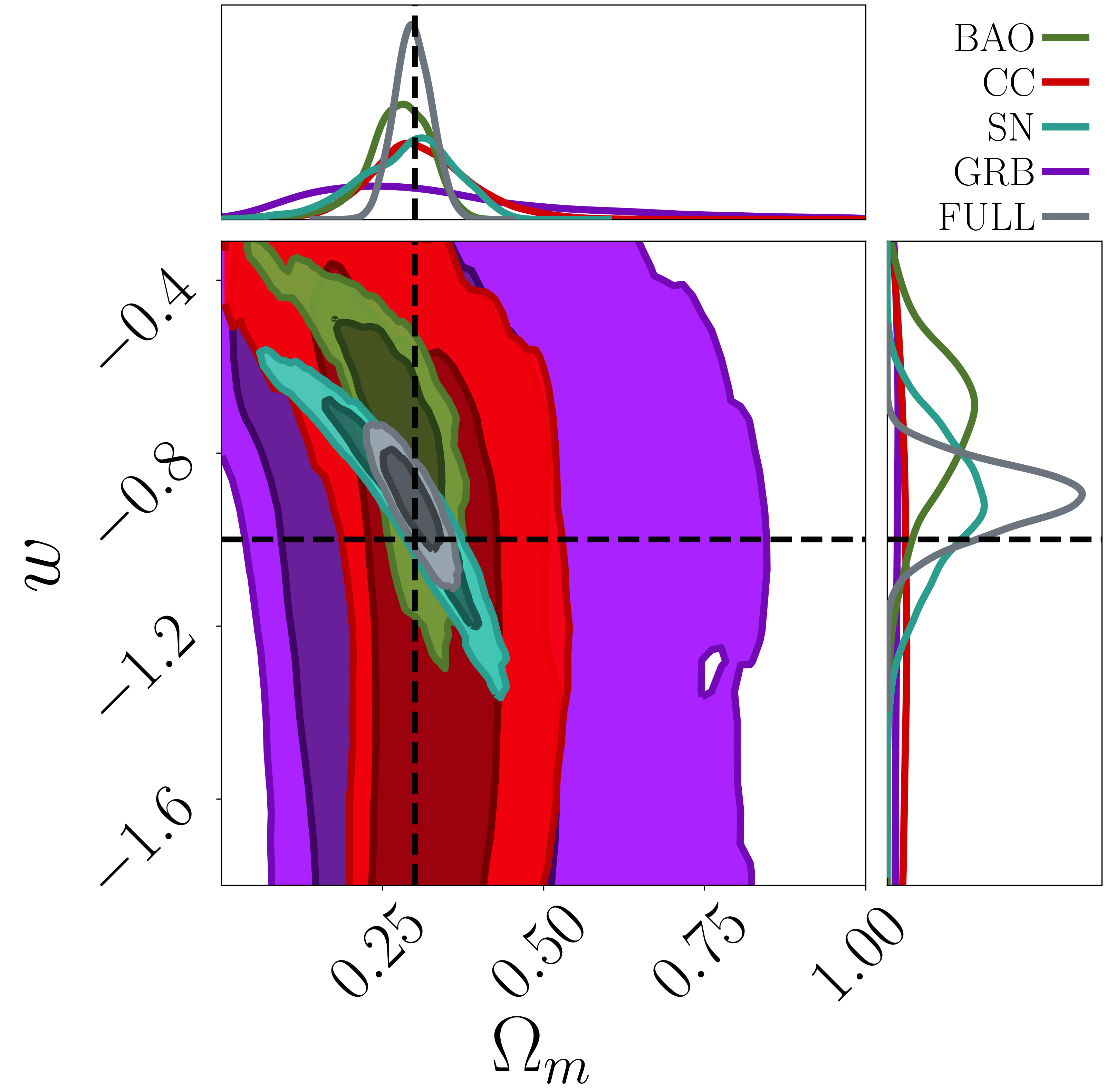}
    \caption{flat $w$CDM}
    \label{fig:resFwCDM}
    \end{subfigure}
    \begin{subfigure}[t]{0.395\textwidth}
    \includegraphics[width=\textwidth]{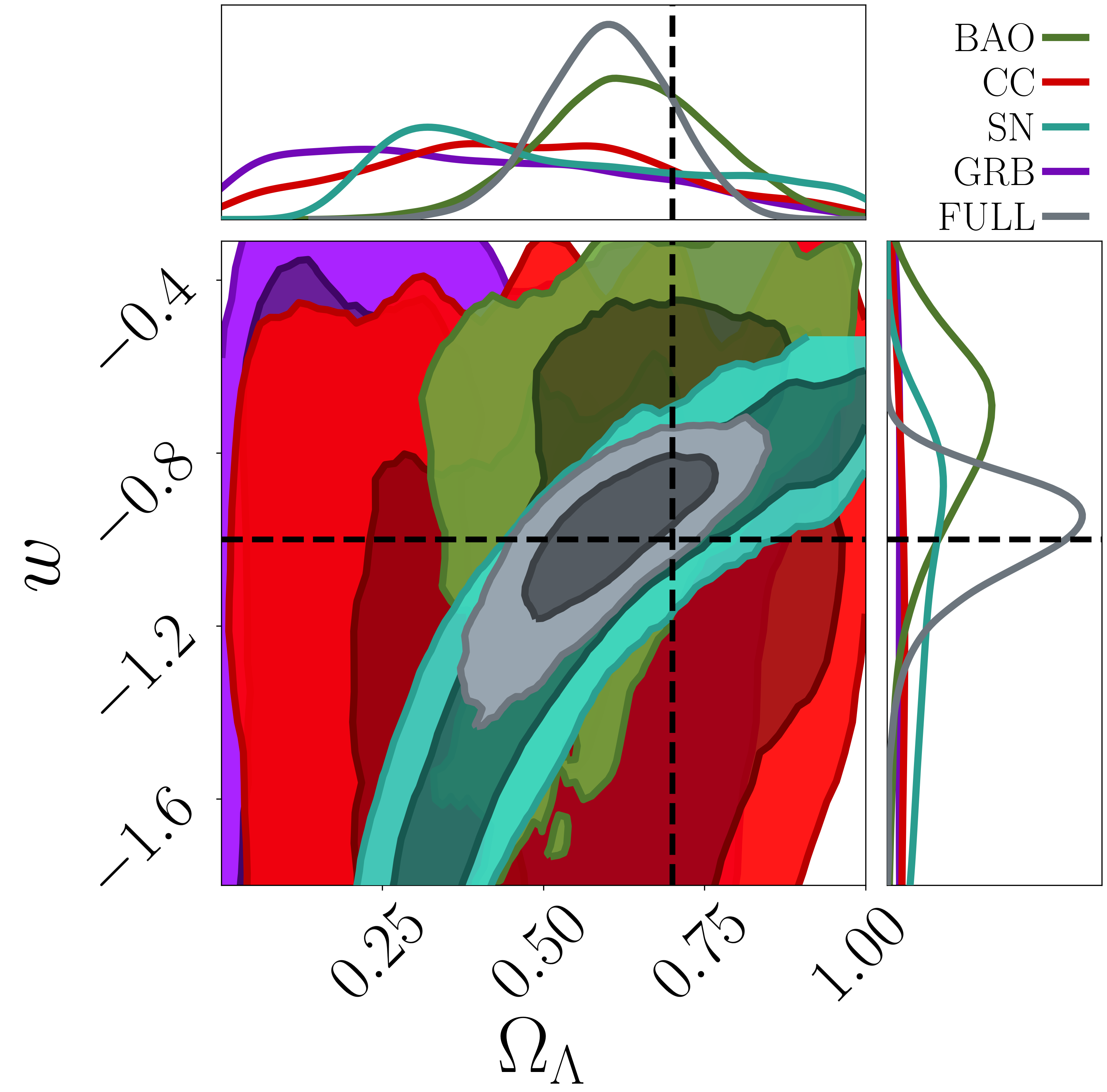}
    \caption{$w$CDM}
    \label{fig:reswCDM}
    \end{subfigure}
    \begin{subfigure}[b]{0.395\textwidth}
    \includegraphics[width=\textwidth]{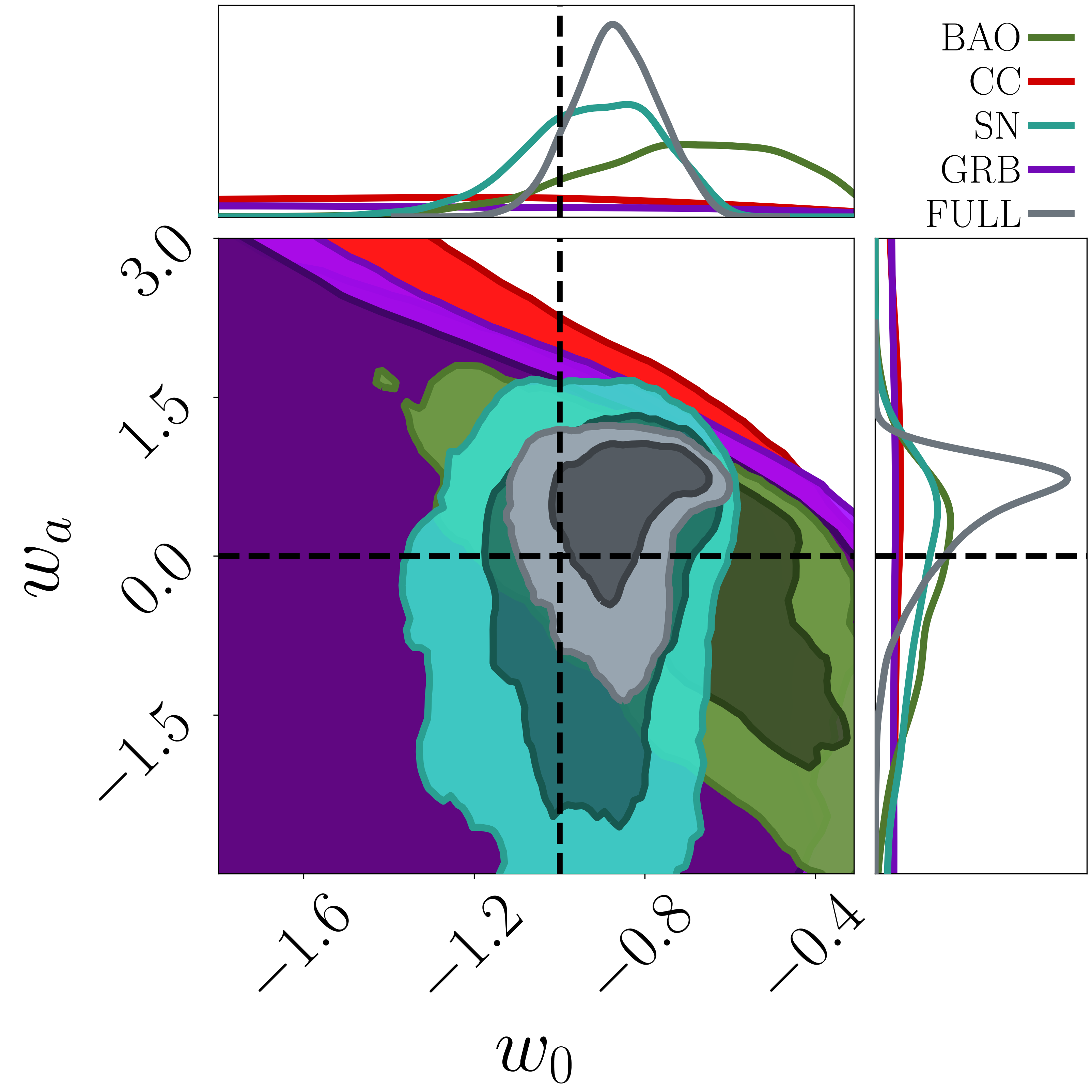}
    \caption{flat $w0wa$CDM}
    \label{fig:resFw0waCDM}
    \end{subfigure}
    \begin{subfigure}[b]{0.395\textwidth}
    \includegraphics[width=\textwidth]{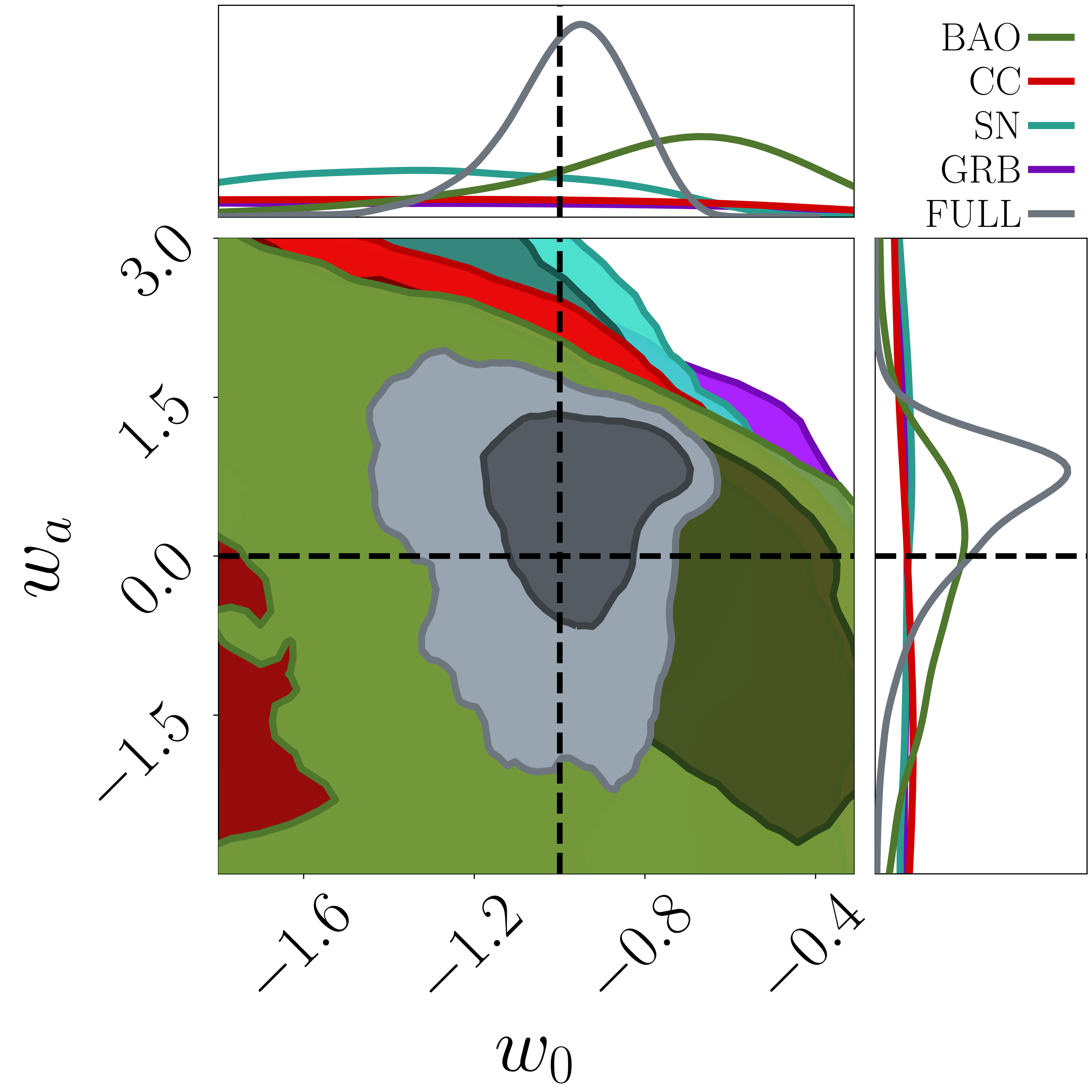}
    \caption{$w0wa$CDM}
    \label{fig:resw0waCDM}
    \end{subfigure}
    \caption{Contour plots (at $68\%$ and $95\%$ C.L.) and one--dimensional marginalized distribution inferred from CCs (red), SNe (cyan) GRBs (violet), BAOs (green) and their full combination (grey). 
    Each panel shows the most significant parameter plane according to the cosmological model.
    Dashed black lines are associated to the reference values: $H_0=70$ km s$^{-1}$ Mpc$^{-1}$, $\Omega_m=0.3$, $\Omega_{\Lambda}=0.7$, $w_0=-1$ and $w_a=0$. 
    }
    \label{fig:res_conclusioni}
\end{figure*}

The likelihood function is constructed through the standard approach:
\begin{equation}
\label{eq:like_TOT}
    	\text{ln}\ \mathcal{L}_{\text{full}}\ = \ \text{ln}\ \mathcal{L}_{\text{BAO}}\ +\  \text{ln}\ \mathcal{L}_{\text{CC}}\ +\ \text{ln}\ \mathcal{L}_{\text{SN}}\ +\ \text{ln}\ \mathcal{L}_{\text{GRB}} 
\end{equation}
where $\text{ln}\mathcal{L}_{\text{BAO}}$, $\text{ln}\mathcal{L}_{\text{SN}}$, $\text{ln}\mathcal{L}_{\text{CC}}$ and $\text{ln}\mathcal{L}_{\text{GRB}}$ are built from Equation \ref{eq:like_BAO}, \ref{eq:like_SN}, \ref{eq:like_CC} and  \ref{eq:like_GRB}, respectively.
The cosmological constraints from the BAO+CC+SN+GRB combination (full, in short) are reported in Table \ref{tab:res_BCSG}, while in Figure \ref{fig:res_conclusioni} we show some specific projections of the sampled posterior distributions.

The combination of these late--Universe probes shows immediately its potential in the modern cosmological context. Assuming a flat $\Lambda$CDM model, we precisely constrain (at 68$\%$ C.L.) the standard cosmological parameters as:
\begin{equation}
\label{eq:FLCDM_res}
\text{\big[BAO+CC+SN+GRB\big]}
    \begin{cases}
    H_0=67.2 \pm 3.3 \ \ \text{km s$^{-1}$ Mpc$^{-1}$} \\ 
    \Omega_m=0.325\pm0.015 
    \end{cases}
\end{equation}
showing remarkable precision on both parameters, especially if we consider that they are obtained without any restrictive prior.

As suggested by Figure \ref{fig:resFLCDM}, CC is the only considered probe sensitive to the Hubble constant. For this reason, it is maybe more interesting to note how by combining CCs with BAOs, SNe, and GRBs we are able to improve the percentage precision on $H_0$ from $8\%$ to $5\%$. On the other hand, the noticeable result achieved for $\Omega_m$ is essentially driven by the high precision of BAOs and SNe constraints.

\begin{table*}
		\begin{center}
			\caption{Constraints on cosmological and nuisance parameters (best--fit and 68$\%$ C.L. values) from the combination of BAOs, CCs, SNe, and GRBs data. Note that $r_d$ and $H_0$ values are respectively in units of Mpc and km s$^{-1}$ Mpc$^{-1}$. Empty values correspond to parameters not constrained by the related cosmological model. The values of $\Omega_k (\equiv 1 - \Omega_m - \Omega_{\Lambda})$ are directly extracted by combining the marginalized distributions of $\Omega_m$ and $\Omega_{\Lambda}$.}
				\label{tab:res_BCSG}
		        \renewcommand{\arraystretch}{1.5}
				\begin{tabular}{ccccccc}
					\hline
				 & flat $\Lambda$CDM & $\Lambda$CDM & flat $w$CDM & $w$CDM & flat $w_0w_a$CDM & $w_0w_a$CDM  \\ 
			    \hline
				\hline	
				\multicolumn{1}{c}{} & \multicolumn{6}{c}{\textbf{BAO+CC+SN+GRB}} \\
				\hline	
                $H_0$ & $67.2^{+3.4}_{-3.2}$ & $67.0^{+3.2}_{-3.4}$ & $67.5^{+3.3}_{-3.4}$ & $66.9^{+3.5}_{-3.4}$ & $67.7^{+3.5}_{-3.3}$ & $66.8^{+3.3}_{-3.3}$ \\ [0.3ex] 
     \hline
     $\Omega_m$ & $0.325^{+0.015}_{-0.015}$ & $0.279^{+0.034}_{-0.033}$ & $0.299^{+0.027}_{-0.028}$ & $0.276^{+0.033}_{-0.032}$ & $0.258^{+0.058}_{-0.101}$ & $0.23^{+0.058}_{-0.091}$ \\ [0.3ex] 
     \hline
     $\Omega_{\Lambda}$ & -- & $0.596^{+0.053}_{-0.055}$ & -- & $0.601^{+0.095}_{-0.102}$ & -- & $0.627^{+0.137}_{-0.124}$ \\ [0.3ex] 
     \hline
     $\Omega_k$ & -- & $0.125^{+0.081}_{-0.083}$ & -- & $0.122^{+0.117}_{-0.109}$ & -- & $0.148^{+0.118}_{-0.111}$ \\ [0.3ex] 
     \hline
     $w_0$ & -- & -- & $-0.91^{+0.07}_{-0.08}$ & $-0.98^{+0.11}_{-0.16}$ & $-0.87^{+0.09}_{-0.09}$ & $-0.96^{+0.14}_{-0.18}$ \\ [0.3ex] 
     \hline
     $w_a$ & -- & -- & -- & -- & $0.47^{+0.36}_{-0.66}$ & $0.61^{+0.42}_{-0.83}$ \\ [0.3ex] 
     \hline
     $M$ & $-19.4^{+0.1}_{-0.1}$ & $-19.4^{+0.1}_{-0.1}$ & $-19.4^{+0.1}_{-0.1}$ & $-19.4^{+0.1}_{-0.1}$ & $-19.4^{+0.1}_{-0.1}$ & $-19.4^{+0.1}_{-0.1}$ \\ [0.3ex] 
     \hline
     $a$ & $0.479^{+0.018}_{-0.018}$ & $0.475^{+0.019}_{-0.018}$ & $0.478^{+0.019}_{-0.019}$ & $0.475^{+0.019}_{-0.018}$ & $0.479^{+0.017}_{-0.018}$ & $0.475^{+0.018}_{-0.019}$ \\ [0.3ex] 
     \hline
     $b$ & $2.05^{+0.03}_{-0.03}$ & $2.05^{+0.03}_{-0.03}$ & $2.06^{+0.03}_{-0.03}$ & $2.05^{+0.03}_{-0.03}$ & $2.06^{+0.03}_{-0.03}$ & $2.05^{+0.03}_{-0.03}$ \\ [0.3ex] 
     \hline
     $\sigma_{int}$ & $0.213^{+0.011}_{-0.011}$ & $0.213^{+0.012}_{-0.011}$ & $0.213^{+0.012}_{-0.011}$ & $0.212^{+0.011}_{-0.011}$ & $0.213^{+0.012}_{-0.011}$ & $0.213^{+0.011}_{-0.011}$ \\ [0.3ex] 
     \hline
     $r_d$ & $147.7^{+7.3}_{-7.0}$ & $147.3^{+7.6}_{-6.6}$ & $146.3^{+7.7}_{-6.8}$ & $147.3^{+7.8}_{-7.1}$ & $146.1^{+7.4}_{-7.2}$ & $147.6^{+7.5}_{-7.1}$ \\ [0.3ex] 
				\hline
				\hline
				\end{tabular}
			\end{center}
\end{table*}

We note also the tendency of the BAO+CC+SN+GRB combination to prefer a Universe with negative curvature ($k=-1\Rightarrow\Omega_k>0$), but this trend is not significant at $95\%$ C.L, and therefore does not represent a statistically significant deviation from the standard paradigm.
Finally, considering the most complicated model, i.e., $w_0w_a$CDM, the DE EoS seems to be consistent with a Cosmological Constant $\Lambda$, even though Figure \ref{fig:resw0waCDM} demonstrates how the one--dimensional marginalized distribution of $w_a$ appears particularly asymmetric, with the peak of the distribution around $w_a\sim1$.

\subsubsection{Comparison with literature results}
It is very interesting to compare our results to the ones in the literature considering other (combinations of) probes.

First of all, our measurement of the Hubble constant in the flat $\Lambda$CDM cosmology is fully consistent with the early--Universe (TT,TE,EE+lowE+lensing) inference derived by \citet{planck18}:
\begin{equation}
    \text{\big[CMB\big]}\ \ \ \ \ H_0=67.36\pm0.54 \ \text{km s$^{-1}$ Mpc$^{-1}$  \ (68\%\ C.L.)} 
\end{equation}
and, instead, slightly deviates at 68$\%$ C.L. from the latest value found by \citet{shoes_pantheon+} from the combination of local Cepheid variables and Pantheon+ SNe:
\begin{equation}
    \text{\big[SH0ES\big]}\ \ \ \ \ H_0=73.6\pm1.1 \ \text{km s$^{-1}$ Mpc$^{-1}$  \ (68\%\ C.L.)} \ .
\end{equation}
Although we do not achieve enough precision to disentangle between the $H_0$ current values of the Hubble tension, we note that our results prefer solutions leaning towards a CMB--like measurement. Moreover, they show remarkable stability (in terms of precision $\delta H_0$) as the model dimension increases. 

Since it is known the strong model dependence of the constraints from CMB power spectrum (TT,TE,EE+lowE), we also compare our $H_0$ measurement to a cosmological independent result, such as the combination between CMB lensing and BAOs \citep{planck18}. With a prior on the baryon density ($\Omega_b h^2 = 0.0222 \pm 0.0005$), the standard cosmological model is constrained (68\% C.L.) as:
\begin{equation}
    \text{\big[CMB lensing+BAO\big]}\ \
    \begin{cases}
     H_0=67.9\pm1.3 \  \text{km s$^{-1}$ Mpc$^{-1}$} \\
     \Omega_m=0.303\pm0.017
    \end{cases}
\end{equation}
with respect to which our late--Universe analysis, without applying any restrictive prior, is extremely competitive for what concerns both $H_0$ and $\Omega_m$.

By further looking at the literature results, it clearly emerges how our approach represents a worthy rival to the standard probes usually combined in modern cosmology.
In fact, by assuming a flat $w$CDM model and combining CMB (TT,TE,EE+lowE) with some BAOs data -- including \citet{Ross15}, \citet{Alam17}, \citet{dumas21} and \citet{eBOSSDR16} -- and the Pantheon+ sample, \citet{shoes_pantheon+} have inferred (at 68$\%$ C.L.):
\begin{equation}
    \text{\big[CMB+BAO+SN(Pantheon+)\big]}\ \ 
    \begin{cases}
     \Omega_m=0.316\ ^{+0.005}_{-0.008} \\
     w_0=-0.978^{+0.024}_{-0.031}
    \end{cases}
\end{equation}
while, by excluding BAOs data, they found:
\begin{equation}
    \text{\big[CMB+SN(Pantheon+)\big]}\ \ 
    \begin{cases}
     \Omega_m=0.325\ ^{+0.010}_{-0.008} \\
     w_0=-0.982^{+0.022}_{-0.038}
    \end{cases}
\end{equation}
with a precision on both parameters about 2 times greater than those achieved in this work.
\begin{figure*}
    \centering
    \begin{subfigure}[t]{0.39\textwidth}
    \includegraphics[width=\textwidth]{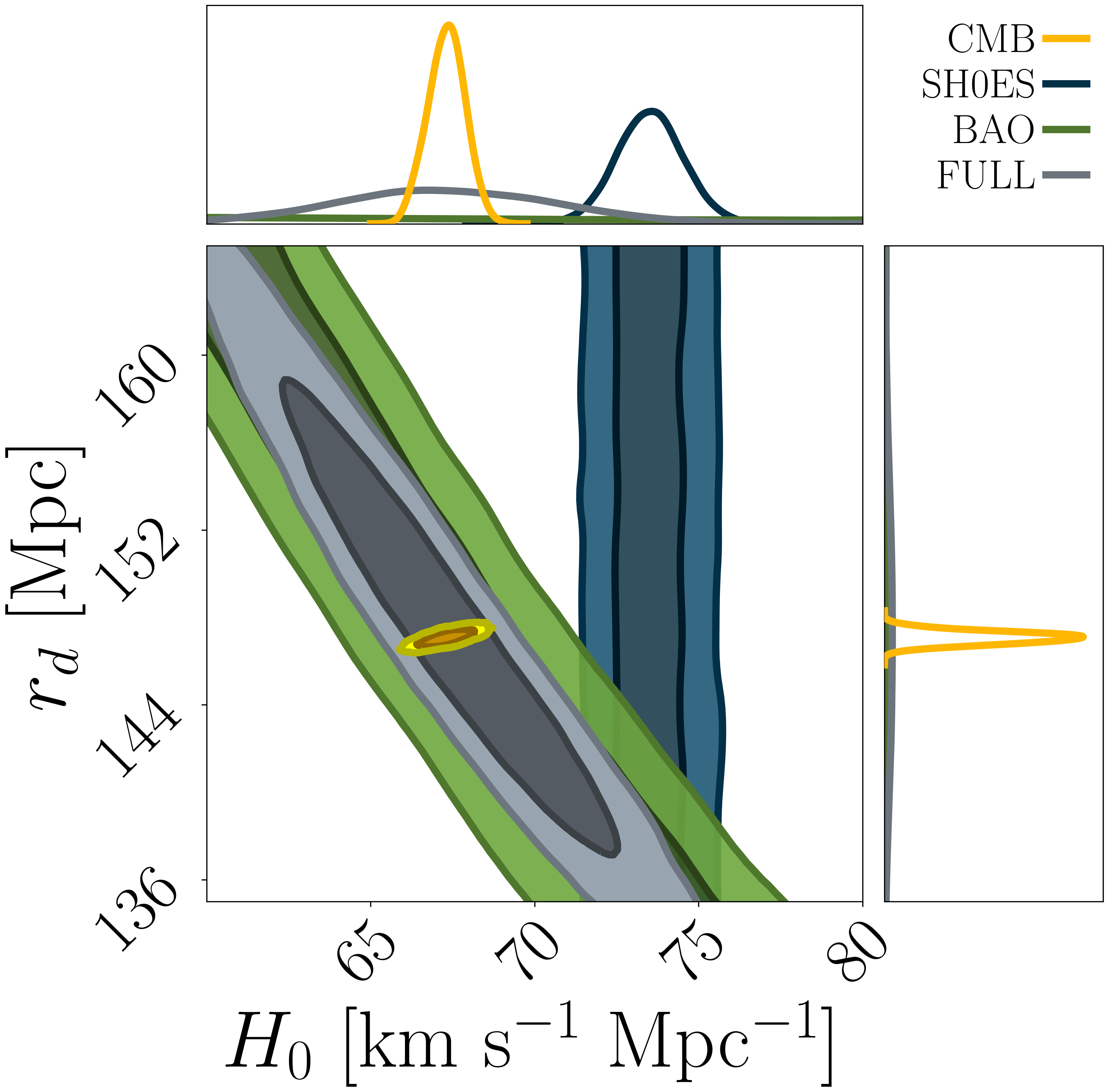}
    \caption{$H_0-r_d$ plane}
    \label{fig:H0rd_FLCDM}
    \end{subfigure}
    \begin{subfigure}[t]{0.402\textwidth}
    \includegraphics[width=\textwidth]{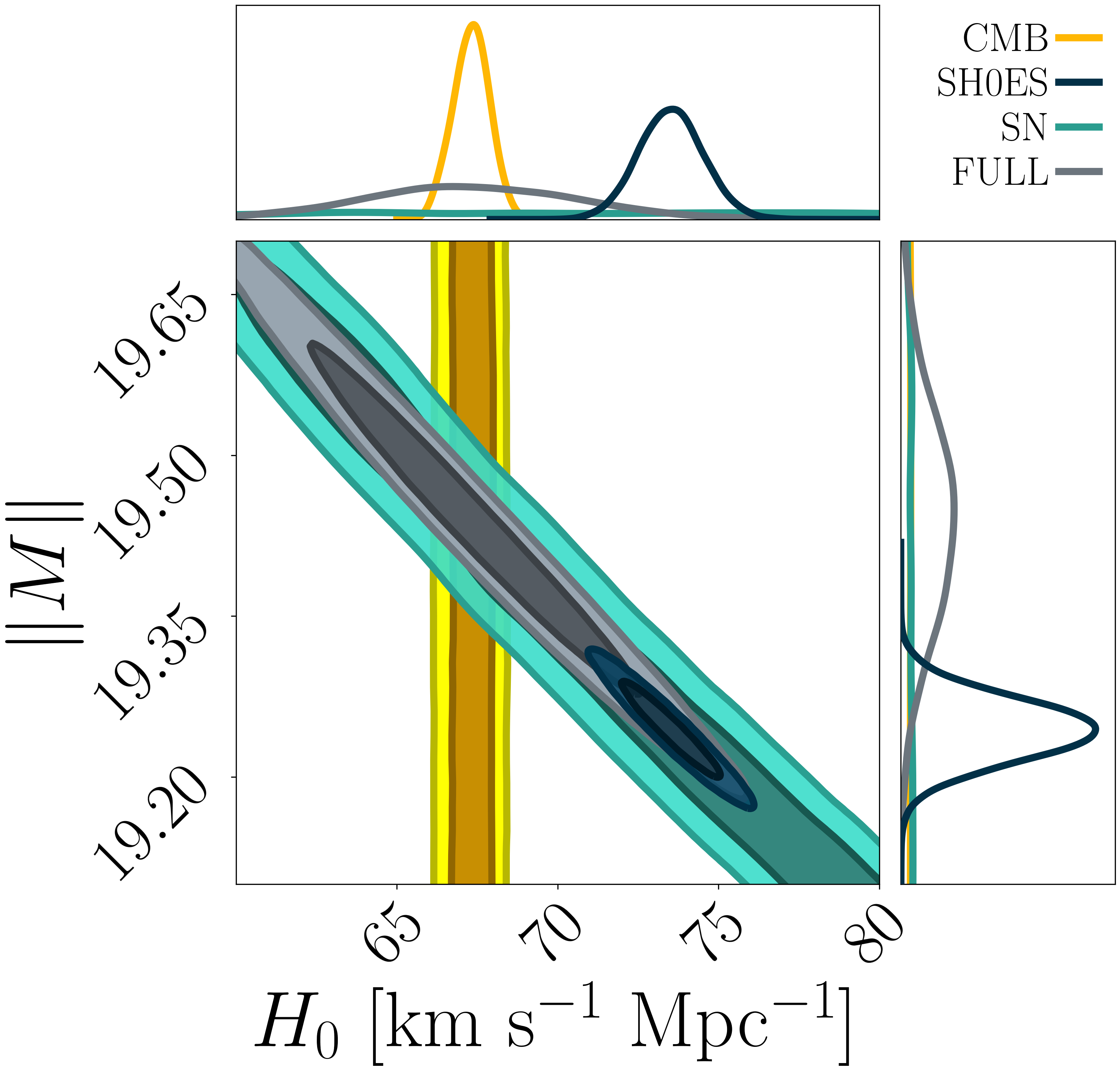}
    \caption{$H_0-M$ plane}
    \label{fig:H0M_FLCDM}
    \end{subfigure}
    \caption{Overlap of the two--dimensional contours (at $68\%$ and $95\%$ C.L.) of different probes in a flat $\Lambda$CDM model. We show CMB \citep[yellow]{planck18}, SH0ES \citep[blue]{shoes_pantheon+}, BAOs (green), SNe (cyan) and full (grey) contours in the $H_0-r_d$ (left) and $H_0-M$ (right) planes. Note that $M$ contours are shown in absolute value.}
    \label{fig:nuisance}
\end{figure*}
In addition, by analysing the same combinations in the case of an evolutionary DE EoS (flat $w_0w_a$CDM), they found the following constraints: 
\begin{equation}
    \text{\big[CMB+BAO+SN(Pantheon+)\big]}\ \ 
    \begin{cases}
     \Omega_m=0.318\ ^{+0.009}_{-0.005} \\
     w_0=-0.841^{+0.066}_{-0.061} \\
     w_a=-0.65^{+0.28}_{-0.32} \\
    \end{cases}
\end{equation}
\begin{equation}
    \text{\big[CMB+SN(Pantheon+)\big]}\ \ 
    \begin{cases}
    \Omega_m=0.318\ ^{+0.012}_{-0.014} \\
     w_0=-0.851^{+0.092}_{-0.099} \\
     w_a=-0.70^{+0.49}_{-0.51} \\
    \end{cases}
\end{equation}
In all these cases, the CMB power spectrum acts as the main driver of the high precision achieved. 
Especially for high dimensional models, it is hence essential to emphasize once again the extreme relevance of a combination of late-Universe probes, since this method has proven its reliability in providing cosmological results that are competitive with those from standard cosmological literature.
Such a relevant outcome is essentially driven by the different cosmological dependencies depicted by each probe selected for this work. 
As shown in Figure \ref{fig:res_conclusioni}, according to which observable and empirical relationship they make use of to constrain cosmological parameters, distinct probes show different oriented contours. So, the combination of many different probes can exploit several constraining powers in order to break (or, at least, to alleviate) some of the parametric degenerations characterizing every cosmological model.
The striking examples of this behaviour are observed in Figure \ref{fig:resFwCDM} and \ref{fig:resFw0waCDM} where the pronounced orthogonality between BAOs and SNe (supported also by the contribution of CCs and GRBs constraints) significantly increases the precision on the cosmological parameter estimation.

All these considerations are underlying the fundamental role played by probes combination in the modern cosmological context and its innovative potential to address some of the still open questions. As we will study and deeply analyse in Section \ref{sec:syn_com}, in a scenario dominated by strong parametric degenerations it is worth implementing the combination of probes as a tool to break them and improve the inference of cosmological parameters.

\subsubsection{A nuisance parameters' perspective on the Hubble tension}
\label{sec:nuisance_vs_H0}
In this work, in addition to the cosmological ones, we also constrain some nuisance parameters, such as the absolute magnitude $M$ of a fiducial SNe, the radius $r_d$ of the sound horizon evaluated at the decoupling epoch, as well as the slope $a$ and the intercept $b$ of the ``Amati'' relation.
The dependence of the observables on these quantities introduces a degeneracy with the Hubble constant. 
For this reason, it is possible to constrain $H_0$ only when SNe, BAOs or GRBs are combined with $H_0$--sensitive probes, such as the CCs \citep{morescorev22}. 
Another way to break the degeneracy is to use an external calibrator able to constrain the nuisance parameters, namely, the observation of local Cepheid variables \citep{shoes_pantheon+} for calibrating the absolute magnitude $M$ of SNe, or the CMB power spectrum \citep{planck18} to directly measure the length of $r_d$.

Starting from this consideration, we derive two important consequences.
First of all, including CC data allows us to obtain more precise measurements of nuisance parameters and to calibrate the SNe, BAOs and GRBs methods. The effects of the CC--calibration on other probes can be seen by observing how the precision on nuisance parameters increases passing from the individual probes analysis (Table \ref{tab:res_conclusioni}) to the full combination one (Table \ref{tab:res_BCSG}). 
At the same time, considering the full probes combination, we notice that the inferred precision on $H_0$ is almost constant, contrary to what would be expected as the dimension of the analysed cosmological model increases. 
It is interesting to notice (looking at Table \ref{tab:res_conclusioni}) that similar behaviour is observed also for nuisance parameters from individual probe analyses (i.e., $M$ for SNe, $r_d$ for BAOs, $b$ for GRBs).
This effect points to the fact that these nuisance parameters do not depend on the cosmological model, but only on observational data. Therefore, a possible explanation would be that in the case of the full combination of probes, the uncertainties on $H_0$ cannot be reduced below a given threshold given by the intrinsic scatter in the SNe, BAO and GRB data, representing a plateau in the $H_0$ error (see Table \ref{tab:res_BCSG}).

Secondly, as already established by several works, such as \citet{efstathiou}, \citet{Alam21} or \citet{favale23}, the difference between values of the current expansion rate of the Universe found by \citet{planck18} and \citet{shoes} may also be read as a discrepancy between the inferred values of the nuisance parameters of the BAOs and SNe methods, i.e., $r_d$ and $M$, respectively. From this perspective, Figure \ref{fig:nuisance} shows how our combination of late--Universe probes seems to prefer values of $r_d$ and $M$ consistent with those obtained with the inverse distance ladder, i.e., through a calibration from CMB data \citep{planck18}, rather than those found from the three--rung distance ladder in the local Universe \citep{shoes_pantheon+}.
Indeed, considering a flat $\Lambda$CDM model, from the full combination we measured these two parameters as:
\begin{equation}
    \text{\big[BAO+CC+SN+GRB\big]}\ \ 
    \begin{cases}
     r_d =  148 \pm 7 \ \text{Mpc} \\
     M = - 19.4 \pm 0.1 \ \text{mag}
    \end{cases}
\end{equation}
showing a small deviation to results obtained by \citet{Alam21}:
\begin{equation}
     r_d =  136 \pm 3 \ \text{Mpc}
\end{equation}
and by \citet{shoes_pantheon+}:
\begin{equation}
     M =  - 19.24 \pm 0.03 \ \text{mag}
\end{equation}
from the calibration of the sound horizon $r_d$ and the absolute magnitude $M$ through the three-rung distance ladder.

\section{Synergies and complementarities}
\label{sec:syn_com}
We conclude our analysis with an analytical quantification of the contributions of different datasets to the inference of cosmological parameters.
Our intention is to figure out the role of each probe in the improvement of precision of the various parameters, quantify the complementarities between them, and study how to exploit synergies in order to maximize the constraint powers of cosmological data.

\subsection{Degeneracy and orthogonality}
\label{sec:com}
As implicitly assumed, every dataset has a peculiar constraining power reflected in different orientations of the contours that each probe shows in specific planes (see Figure \ref{fig:res_conclusioni}).

Here, we define a method to describe the geometric properties of contours within the parameter space.
The first step is the determination of the degeneracy direction, i.e., the orientation of the contours on a defined parameter plane. For this purpose, we extract from the MCMC samples the marginalised 2D distribution of the posterior for a given combination of cosmological parameters and then estimate the counter--clockwise angle $\beta$ of the contours as the inclination of the major axis of the ellipse defined by the eigenvalues and eigenvectors of the 2D contour.
We then define the degeneracy direction as the line -- passing through the $50-$th percentile of the distribution -- with slope:
\begin{equation}
    m\ \equiv\ \text{tan}\ \big(\beta\big)\ .
\end{equation}
It is then straightforward to determine the relative orientation of the contours of different probes by considering the acute angle $\alpha$ between two different degeneracy directions. 
As a result, geometrically speaking, we quantify the orthogonality factor $\varkappa$ between two probes by defining:
\begin{equation}
    \varkappa\ \equiv \ \frac{2\alpha}{\pi}
\end{equation}
so that two perfectly orthogonal (parallel) contours are characterized by $\varkappa = 1$ ($\varkappa = 0$).

Because of the different ranges spanned by the various cosmological parameters (see Table \ref{tab:prior_std}), when calculating the relative orientation $\alpha$, we normalize the sampled distributions to one to avoid scale bias and improve the accuracy on $\varkappa$. In practice, while we do nothing for the energy density parameters, we renormalise the range of $H_0$ and $w_0$ to one (equivalently to considering $h$). 
Specifically:
\begin{equation}
    d_{\text{Norm}}\ \equiv\ \frac{d_{\text{MC}} - p_{\text{min}}}{p_{\text{max}} - p_{\text{min}}}
\end{equation}
where $d_{\text{MC}}$ is the original MCMC distribution, $p_{\text{min}}$ and $p_{\text{max}}$ are the extremes of the prior intervals.

\begin{figure}
    \centering
    \includegraphics[width=0.47\textwidth]{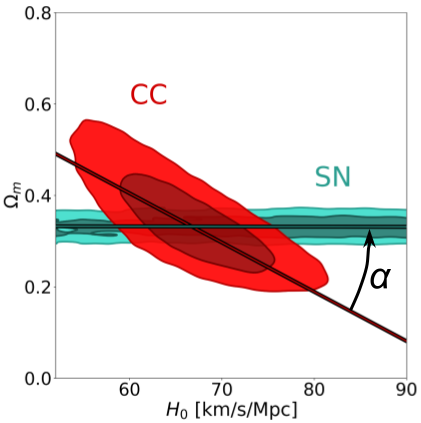}
    \caption{Example of the calculation of degeneracy directions and relative orientation $\alpha$. The contours obtained from the CCs and SNe analyses of the flat $\Lambda$CDM model are shown in the $H_0-\Omega_m$ plane.}
    \label{fig:deg}
\end{figure}

As an example, in Figure \ref{fig:deg}, the relative orientation $\alpha$ of the SNe and CCs contours is shown in the $H_0-\Omega_m$ plane for the flat $\Lambda$CDM model. 
As reported in Table \ref{tab:ort}, these two probes have an orthogonality factor $\varkappa = 0.35$ corresponding to a relative orientation $\alpha = 31.4^{\circ}$.

Obviously, because the accuracy of this estimate strongly depends on the calculation of the covariance ellipse from the MCMC sampling, the method implemented here is not sensitive to distributions significantly deviating from the multivariate Gaussian. 
For this reason, in Table \ref{tab:ort} we show only those parameter planes in which contours are approximately ellipsoidal and not extremely widespread, i.e., flat $\Lambda$CDM, $\Lambda$CDM and flat $w$CDM (see Figure \ref{fig:res_conclusioni}).
\begin{table} 
     \renewcommand{\arraystretch}{1.5}
     \caption{Relative orientation $\alpha$ and orthogonality $\varkappa$ between BAOs, CCs, SN and GRBs on particular parameter planes within the parameter space.}
	 \label{tab:ort}
     \begin{center}
     \begin{tabular}{c||c|c||c|c||c|c}
     \hline
       & \multicolumn{2}{c||}{flat $\Lambda$CDM} & \multicolumn{2}{c||}{$\Lambda$CDM} & \multicolumn{2}{c}{flat $w$CDM} \\ 
       & \multicolumn{2}{c||}{$H_0-\Omega_m$} & \multicolumn{2}{c||}{$\Omega_m-\Omega_{\Lambda}$} & \multicolumn{2}{c}{$\Omega_m-w$} \\ 
     \hline
      & $\alpha$ & $\varkappa$ & $\alpha$ & $\varkappa$ & $\alpha$ & $\varkappa$ \\[0.3ex] 
     \hline
     \hline
     $BA$$\widehat{O \ C}$$C$ & $31.3^\circ$ & $0.35$ & $7.7^\circ$ & $0.09$ & $17.8^\circ$ & $0.20$  \\ [0.3ex]  
     $BA$$\widehat{O \ S}$$N$ & $0.1^\circ$ & $0.00$ & $13.5^\circ$ & $0.15$ & $2.5^\circ$ & $0.03$ \\ [0.3ex]  
     $BA$$\widehat{O \ G}$$RB$ & $0.7^\circ$ & $0.01$ & $36.7^\circ$ & $0.41$ & $12.7^\circ$ & $0.14$ \\ [0.3ex]  
     $C$$\widehat{C \ S}$$N$ & $31.4^\circ$ & $0.35$ & $5.8^\circ$ & $0.06$ & $20.3^\circ$ & $0.23$ \\ [0.3ex]  
     $C$$\widehat{C \ G}$$RB$ & $30.6^\circ$ & $0.34$ & $44.4^\circ$ & $0.49$ & $30.6^\circ$ & $0.34$ \\ [0.3ex]  
     $S$$\widehat{N \ G}$$RB$ & $0.8^\circ$ & $0.01$ & $50.2^\circ$ & $0.56$ & $10.2^\circ$ & $0.11$ \\ [0.3ex]  
     \hline
     \hline
     \end{tabular}
     \end{center}
\end{table}
\begin{table*} 
\renewcommand{\arraystretch}{1.5}
     \begin{center}
     \caption{Figure--of--Merit obtained for particular planes within the parameter space by BAOs, CCs, SNe, GRBs and their combination BAO+CC+SN+GRB.}
     \label{tab:FoM}
     \begin{tabular}{c||c|c|c|c|c|c}
     \hline
     \multicolumn{7}{c}{\textbf{Figure--of--Merit}} \\
     \hline
      & \ \ \ \ flat $\Lambda$CDM\ \ \ \ &\ \ \ \ $\Lambda$CDM\ \ \ \ &\ \ \ \  flat $w$CDM\ \ \ \ &\ \ \ \ $w$CDM\ \ \ \ & flat $w_0w_a$CDM & $w_0w_a$CDM  \\
      & $H_0 - \Omega_m$ & $\Omega_m - \Omega_{\Lambda}$ & $\Omega_m - w_0$ & $\Omega_{\Lambda} - w_0$ & $w_0 - w_a$ & $w_0 - w_a$  \\ 
     \hline
     \hline
     BAO+CC+SN+GRB & 20.8 & 827.6 & 890.0 & 118.8 & 19.1 & 6.8 \\ [0.3ex] 
     \hline
     BAO & 2.0 & 264.9 & 142.2 & 21.8 & 5.4 & 1.1 \\ [0.3ex]  
     CC & 4.1 & 35.1 & 14.2 & 3.6 & 0.5 & 0.3  \\ [0.3ex]  
     SN & 3.9 & 428.8 & 253.3 & 13.5 & 4.3 & 0.6  \\ [0.3ex]  
     GRB & 0.2 & 25.6 & 3.6 & 3.0 & 0.3 & 0.4   \\ [0.3ex] 
     \hline
     \hline
     \end{tabular}
     \end{center}
\end{table*}
\subsection{Figure--of--Merit}
\label{sec:syn}
Since the evidence of a new component emerged in the cosmological energy budget, the scientific community has developed several tools to quantify how future experiments could improve the measurement of Dark Energy properties \citep{linder06, albrecht06, albrecht07, yunwangFoM, sartoris16}.
\\
Considering a cosmological model $\mathcal{C}$ described by a set of N parameters $f_{\mathcal{C}} \equiv \{f_1, f_2, ... , f_N\}$, the Figure--of--Merit (FoM) associated to the constraints provided by a given probe is defined as:
\begin{equation}
\label{eq:FoMDE}
    \text{FoM}_{\mathcal{C}} \equiv \frac{1}{\sqrt{\text{det\ Cov}(f_{\mathcal{C}})}}
\end{equation}
where $Cov$ is the covariance matrix between the measured $f_{\mathcal{C}}$ parameters.
Generally speaking, if the posterior is a multivariate Gaussian distribution, the volume that it occupies within the entire cosmological parameter space is proportional to the inverse of the FoM.
As suggested by \citet{linder06}, the volume of the probability distribution represents a suitable way to evaluate how different probes constrain the parameter space. But we shall keep in mind that it can be used as a proxy of the precision achieved by a chosen dataset only in the Gaussian approximation, i.e., as long as the posterior distribution is a multivariate Gaussian. 

Here, we focus on the main set of cosmological parameters relevant to our analysis. In particular, as already done for the contours shown in Figure \ref{fig:res_conclusioni}, the results obtained for those specific parameters planes are shown in Table \ref{tab:FoM}.

The most relevant results we obtained are the following:
\begin{itemize}
    \item[1.] Through this approach, we quantified the gain in terms of FoM that the full combination achieves w.r.t. the individual probes. Indeed, the FoM values retrieved by the BAO+CC+SN+GRB inference are systematically higher than the sum of FoM from the individual datasets, with a maximum gain of about 3 times for the flat $w$CDM and $w_0w_a$CDM cases.
    \item[2.] The capability of CCs to measure $H_0$ makes them the most powerful probe (among the one considered) in the $H_0-\Omega_m$ plane (flat $\Lambda$CDM), even if the constraint on $\Omega_m$ is weaker w.r.t. the ones from BAOs and SNe (see Table \ref{tab:res_conclusioni}).
    Although SNe are not able to constrain $H_0$, the FoM of SNe and CCs are essentially the same, emphasizing the fundamental contribution that SNe provide for those analyses aiming to measure $\Omega_m$ precisely.
    \item[3.] BAOs and SNe are undoubtedly the most powerful probes that one should exploit whatever the cosmological model considered. While in the flat $\Lambda$CDM, $\Lambda$CDM and flat $w$CDM models, SNe represent by far the most constraining probe; in the other models, the situation is reversed and BAOs achieve higher precision.
    In terms of contribution to the FoM, this makes the BAO+SN combination dominant on all parameter planes, but flat $\Lambda$CDM. For example, in the $\Lambda$CDM case, this combination reaches around $90\%$ of the FoM achieved by BAO+CC+SN+GRB.
    \item[4.] The contribution of GRBs to the FoM in the $H_0-\Omega_m$ plan (flat $\Lambda$CDM) is minimal. This can also be confirmed by looking at the FoM obtained with the BAO+CC+SN combination, which is roughly the same as the full combination.
    Conversely, their contribution increases for the other models, with a maximum of 25.6 in the $\Omega_m-\Omega_{\Lambda}$ plane ($\Lambda$CDM). In fact, the wider the range of redshift sampled by the probe, the more sensitive the probe is to changes in density and DE EoS parameters, remarking how the extension in redshift of GRBs observations (see Figure \ref{fig:dist_data}) is fundamental to better constrain both the energy budget of the Universe and the Dark Energy properties.
\end{itemize}    

However, we emphasize that the Gaussian approximation hypothesis is not consistent with the more complicated models ($w$CDM, flat $w_0w_a$CDM and $w_0w_a$CDM) where one--dimensional marginal distributions begin to be strongly asymmetric (see Figure \ref{fig:res_conclusioni}). 
Therefore, for these models, the FoM values reported in Table \ref{tab:FoM} should be taken as an indication of the relative gain of the full combination, but they are not informative of the volume occupied by the posteriors in the parameter space.

\section{Conclusions}
\label{sec:conclusions}
In this work, we demonstrated how the combination of some late--Universe cosmological probes, such as Baryon Acoustic Oscillations (BAOs), Cosmic Chronometers (CCs), Type Ia Supernovae (SNe), and Gamma--Ray Bursts (GRBs), could be profitably exploited in modern cosmological analysis to obtain precise constraints completely independent of early--Universe or local anchors, such as the Cosmic Microwave Background or the three-rung distance ladder.

The main results can be summarized as follows.
\begin{itemize}
\item Firstly, we developed, tested and validated a Bayesian code to derive the posterior distribution of a variety of cosmological parameters (considering different cosmological models of increasing complexity) considering several different probes. Once we demonstrated the reliability of our analysis by reproducing the main results coming from the literature for each one of the individual probes, we applied a probe--by--probe combination procedure aiming at extracting the cosmological signal from any combination achievable with our selection of probes. This approach has been useful to ensure that no remarkable biases and no systematic effects were affecting the Bayesian analyses.
\item Surely, BAOs and SNe, as the most robust and widely studied probes, are found to contribute significantly to most of the results. 
For example, in the $\Lambda$CDM case, they are respectively found to be responsible for $\sim30\%$ and $\sim50\%$ of the FoM and reach around $90\%$ when combined together.
At the same time, they are restricted to a relatively small redshift range ($z<1.5$) and, if uncalibrated, they are not sensitive to the Hubble constant $H_0$.
From this perspective, it is not surprising the need to combine additional probes and better capitalise different sensitivities to cosmological parameters in order to break degeneracies and go beyond individual constraint powers. In this study, CCs and GRBs are the new emerging probes strongly contributing to thinning out that fog not allowing to see possible deviations to the standard cosmology.
\item CCs offer a cosmology--independent constraint on $H_0$, $\Omega_m$, and curvature, even if with a smaller FoM compared to BAOs and SNe. It is however interesting to underline that their constraints allowed us to break the degeneracies between the Hubble constant and some of the nuisance parameters of other probes, like $M$ for SNe and $r_d$ for BAOs, and enhance the constraints on all of them. For example, in a flat $\Lambda$CDM model, by including CCs in the combination of BAOs and SNe, the uncertainty on $H_0$ decreases from about 5.5 (CC--only) to about 3.5 km/s/Mpc (BAO+CC+SN) while the determination of $\Omega_m$ improves from $0.330^{+0.078}_{-0.061}$ to $0.326^{+0.015}_{-0.014}$.
\item The wide redshift range covered by GRBs ($0<z<10$) has been pivotal to improving the constraints on the energy composition of the Universe and the Dark Energy Equation of State, as these quantities are proven to be better measured by having a larger redshift leverage. In particular, adding GRBs to the combination of the other probes enhances the constraints making the posterior distributions more Gaussian. For example, considering a flat $w$CDM model, while BAO+CC+SN measures $w_0=-0.906^{+0.068}_{-0.084}$, the full combination provides $w_0=-0.907^{+0.072}_{-0.077}$.
\item By analysing the full combination of probes (BAO+CC+SN+GRB) that sample an extremely wide range of cosmic redshift ($0<z<10$), we derived interesting and precise constraints on the properties of the Universe.
Without imposing any restrictive priors, some of the main results (at 68 \% C.L.) are:
\begin{itemize}
    \item \small{flat $\Lambda$CDM} \ \ \ \ \ \ \ \ \ \ $H_0=67.2\pm3.3$ \ \ \  \ \ \ \ \ \ \ \ \ $\Omega_m=0.325\pm0.015$
    \\
    \item \small{$\Lambda$CDM} \ \ \ \  \ \ \ \ \ \ \ \ \ \ \ \ \ $\Omega_m=0.279\pm0.034$ \ \ \ \ $\Omega_{\Lambda}=0.596\pm0.055$
    \\
    \item \small{flat $w$CDM}  \ \ \ \ \ \ \ \ \ \ $\Omega_m=0.299\pm0.028$ \ \ \ \ $w_0=-0.91\pm0.08$
    \\
    \item \small{$w$CDM} \ \ \ \ \ \ \ \ \ \ \ \ \ \ \ \ $\Omega_{\Lambda}=0.601\pm0.099$ \ \ \ \ \ $w_0=-0.98\pm0.14$
    \\
    \item \small{flat $w_0w_a$CDM}\ \ \ \ $w_0=-0.87\pm0.09$ \ \ \ \ \ \ \ \ $w_a=0.47\pm0.51$
    \\
    \item \small{$w_0w_a$CDM} \ \ \ \ \ \ \ \ \ $w_0=-0.96\pm0.16$ \ \ \ \ \ \ \ $w_a=0.61\pm0.63$
\end{itemize}
\item By relaxing the assumption of flatness, we obtained remarkable but no statistically significant deviations from the null curvature paradigm. Indeed, such a combination seems always to prefer (at 68\% C.L.) a hyperbolic Universe: $\Omega_k=0.125\pm0.082$ ($\Lambda$CDM), $\Omega_k=0.122\pm0.113$ ($w$CDM) and $\Omega_k=0.148\pm0.115$ ($w_0w_a$CDM).
\item As a matter of fact, current data are not ready \citep[yet, see][]{morescorev22} to provide competitive constraints on $H_0$ to solve the well--known tension.
At the same time, the great contribution to $H_0$ constraints provided by CCs reveals an important feature of modern cosmological analyses.
When combining probes that lack direct constraining power on $H_0$, careful consideration must be given to the impact of nuisance parameters used to calibrate their cosmological observables. Notably, the full combination analysis reveals that the precision on $H_0$ is inherently constrained by the inherent scatter within data derived from probes insensitive to $H_0$, like BAOs, SNe, and GRBs. Consequently, probes like CCs play an essential role in calibrating these $H_0$--insensitive probes, if one wants to explore independent kinds of calibrators. At the same time, precisely and accurately measuring cosmological observables like the Hubble diagram or the Amati relation directly impacts the data scattering, thereby enhancing the precision of $H_0$ measurements obtained through the combination of both sensitive and insensitive probes.
\item For the first time in literature, by introducing the orthogonality factor $\varkappa$ between contours, we quantified the degeneracy directions and the relative orientation of probes on each parameter plane.
Studying the Figure--of--Merit, we provided a clear quantification of the constraint powers of each probe.
\end{itemize}

In conclusion, the precision achieved by combining a limited sample of late--Universe probes has shown extreme competitiveness compared to the combination of probes sampling from both the early-- and late--Universe, such as CMB, BAOs and SNe.
This is extremely important in view of currently ongoing and future large surveys, like Euclid \citep{euclid}, Vera Rubin Observatory \citep{lsst} and Nancy Grace Roman Space Telescope \citep{wfirst}, which data exploitation will further increase the effectiveness of the inference process and, thus, facilitate the transition from precision cosmology to accurate cosmology \citep{peebles, verde_proc}. 

We acknowledge here two possible follow-ups of this work for the next future, that are currently beyond the scope of this paper.
First, the amplitude of matter perturbation (through the parameter $\sigma_8$) could be included amongst the late--Universe analyses by adding the Redshift Space Distortion probe \citep{benisty21, aubert22} to the combination process. 
This step would be fundamental to studying the degeneration of $\sigma_8$ with the other cosmological parameters and could provide an independent perspective to the current tension on $S_8\equiv \sigma_8(\Omega_m/0.3)^{0.5}$ \citep{abdalla22, Poulin23}.
Second, studying other emergent cosmological probes, especially the ones able to measure the Hubble constant independently -- such as Gravitational Waves \citep{abbott17} or Kilonovae \citep{sneppen23}, would certainly be an excellent way to verify the systematics involved in the inference process, and contemporaneously enhance the precision of the estimation of the current expansion rate of the Universe.

\section*{Acknowledgements}
The authors are grateful to the referee for the constructive comments that helped improve the presentation of the paper.
M.M. and A.C. acknowledge the grants ASI n.I/023/12/0 and ASI n.2018-23-HH.0. A.C. acknowledges the support from grant PRIN MIUR 2017 - 20173ML3WW\_001. M.M. acknowledges support from MIUR, PRIN 2017 (grant 20179ZF5KS) and PRIN 2022 (grant XXX2022NY2ZRS\_001).
L.A. acknowledges support by INAF Mini-Grants Programme 2022. 
%%%%%%%%%%%%%%%%%%%%%%%%%%%%%%%%%%%%%%%%%%%%%%%%%%
\section*{Data Availability}
Data \& Code supporting this work are available here: \url{https://github.com/Fabrizio-Cogato/A-late-Universe-approach.git}.

\bibliographystyle{mnras}
\bibliography{A_late_universe_approach_to_the_weaving_of_modern_cosmology}

%%%%%%%%%%%%%%%%%%%%%%%%%%%%%%%%%%%%%%%%%%%%%%%%%%

%%%%%%%%%%%%%%%%%%%%%%%%%%%%%%%%%%%%%%%%%%%%%%%%%%

% Don't change these lines
\bsp	% typesetting comment
\label{lastpage}
\end{document}